\documentclass[DIV=20,10pt,a4paper,twocolumn,headings=normal]{scrartcl}

\usepackage{graphicx}					
\usepackage{amsmath}
\usepackage{amssymb}
\usepackage{bm}		

\usepackage{abstract} 
\usepackage{flushend} 
\usepackage{cuted} 
\usepackage{widetext}
\usepackage{mathenv} 

\usepackage[title]{appendix} 
\usepackage{cite} 

\usepackage[unicode, pdfstartview= FitH]{hyperref} 

\usepackage[english]{babel}

\begin{document}

\renewcommand{\abstractname}{\vspace{-50pt}} 
\renewcommand{\refname}{\vspace{-23pt}}

\title{Statistical mechanics of fluids at a permeable wall}

\author{V. M. Zaskulnikov  \thanks{zaskulnikov@gmail.com}
\\Institute of Chemical Kinetics and Combustion, 
\\Institutskaya, 3,  Novosibirsk, 630090, Russian Federation}

\date{\today}

\twocolumn[

\maketitle

\begin{onecolabstract} 

The problem of surface effects at a fluid boundary created by the force field of finite value is investigated. A classical simple fluid with a locally introduced field imitating a permeable solid is considered. The cases of micro- and macroscopically smooth boundary are examined and the analysis of static membranes is performed within the framework of general consideration. 

Henry constant of adsorption and its connection with Henry constant of absorption, specific surface  $\Omega$- potential (grand potential) $\gamma$ and the surface number density are determined. High-temperature expansions for basic values are obtained. In the low-temperature limit it is shown that the results coincide with the previously considered problem of impermeable wall. 

``The surface tension coefficient'' decomposes into a value proportional to Henry constant of adsorption depending on the position of the separating surface, and a universal nonlinear surface coefficient.

Two approaches to this problem are analyzed: through the surface cluster expansion and through the pressure tensor. 

Within the first approach, the series in powers of activity is obtained for the surface part of the omega- potential. This expression is similar to the cluster expansion for pressure but in contrast to this case the integrals of Ursell factors contain the multipliers depending on the potential of particles interaction with the external field. 

Within the second approach, Kirkwood-Buff formula for $\gamma$ is extended for the case of the systems under consideration (the field of finite value). As a function of activity, depending on the situation, the surface terms may begin with either a linear term or a quadratic one, which corresponds to the presence or absence of adsorption in the classical understanding.  

It is demonstrated that the derivative of the tangential component of pressure tensor with respect to the chemical potential coincides with the near-surface number density (both averaged over the transition region), which, first of all, proves a complete identity of the approaches of ``cluster expansion'' and ``pressure tensor'' within the limits of their domain of existence, and second, gives the near-surface virial expansion which determines the exact equation of state of the ``two-dimensional'' gas of the near- surface region. 

The procedure of the "mechanical definition" of $\gamma$ is upgraded and substantiated. Coincidence of pressure acting on a transverse wall and the tangential component of pressure tensor, both averages over the transition layer as well as the symmetry of the solution with respect to the permutation of sorbent - fluid are demonstrated. 

\vspace{20pt}

\end{onecolabstract}

]

\saythanks

\section{\label{sec:01}Introduction}

Surface effects at the boundary between a fluid and a solid body considered using the methods of statistical mechanics have been a subject of permanent attention of researchers for more than 60 years \cite{Ono1950, hill1959, onocondo1960, hillstatmeh1987, Bellemans1962, Bakri1966, SteckiSokolowski1978, SokolowskiStecki1980, SokolowskiStecki1981, LairdDavidchak2010, IrvingKirkwood1950, Harasima1958, HendersonSwol1984, Navascues1979, NavascuesBerry1977, StillingerBuff1962, SchofieldHenderson1982}. Some earlier works are cited in \cite{onocondo1960, hillstatmeh1987}.

Two approaches to this problem have been formed historically. The first one is based on constructing the surface cluster expansion - a series in powers of the activity for surface values \cite{Ono1950, hill1959, onocondo1960, hillstatmeh1987, Bellemans1962, SokolowskiStecki1980, SokolowskiStecki1981}. The second one appeals to the notion of pressure tensor \cite{KirkwoodBuff1949, Harasima1958, Navascues1979, HendersonSwol1984, SchofieldHenderson1982}.

In those works, a solid body is modeled by a constant field and the structure of the field is such that is has an impenetrable core. 

The reference points of this sphere are ``surface tension coefficient'' (the surface part of the omega potential or the grand potential) $\gamma$ and ``gas adsorption on a solid surface'' (including Henry constant of adsorption). 

Unlike in the case of hard core potential, investigations of a permeable wall are less thorough. As far as we know, works dealing with the statistical mechanical investigation of this problem had been completely absent until the 80-ies of the past century. The works that appeared in this area later use mainly the DFT approach (for example, see \cite{BrykPatrykiejewSokolowski2000}). 

In our previous work \cite{Zaskulnikov201102a} we developed a consistent approach to the problem of the interface of impermeable wall/fluid that allowed us to obtain closed expressions for key values. In addition, the equivalence of ``tension''- and ``adsorption''-based approaches was demonstrated.

The present work spreads the approach of \cite{Zaskulnikov201102a} for the case of a permeable wall. The work gives the general calculation of the omega potential (section \ref{subsec:03a}), specific surface omega potential $\gamma$  (sections \ref{subsec:03b}, \ref{subsec:03e}) and the surface number density (sections \ref{subsec:03c} and \ref{subsec:03e}). The upgraded procedure of the ``mechanical definition'' of $\gamma$ is given (section \ref{subsec:03h}) and Henry constant of adsorption is obtained (section \ref{subsec:03d}). The equivalence of the cluster expansion and the approach based on a pressure tensor is proved (section \ref{subsec:03k}). Finally, the surface virial expansion is given in section \ref{subsec:03p}.

In addition, a connection between Henry constants of adsorption and absorption is established (section \ref{subsec:03d}) and the high-temperature expansion of surface values is presented (section \ref{subsec:03m}). Static membranes are considered in section \ref{subsec:03o}, and the symmetry of the problem in section \ref{subsec:03l}.

\section{\label{sec:02}Primary definitions and relations}

\subsection{\label{subsec:02a}Canonical ensemble}

The probability density to find a given spatial configuration of a specific set of particles \cite[p.181]{hillstatmeh1987} is given by the expression
\begin{equation}
P^{(k)}_{1...k} = \frac{1}{Z_N}\int \limits_V \exp(-\beta U^N_{1...N})d\bm{r}_{k+1}...d\bm{r}_N. 
\label{eq:001}
\end{equation}

Here $N$ is the number of particles in the system,  $\beta = 1/k_BT$, $k_B$ is Boltzmann constant, $T$ is temperature, $U^N_{1...N}$ is the energy of the interaction of particles with each other, $V$ is the volume of the system.  Integration is performed over the coordinates of the particles of the ensemble $\bm{r}_{k+1}...\bm{r}_N$. $Z_N$  is the configuration integral
\begin{equation}
Z_N = \int \limits_V \exp(-\beta U^N_{1...N}) d\bm{r}_1...d\bm{r}_N. 
\label{eq:002}
\end{equation}

Passing to distribution functions for an arbitrary set of particles, we obtain:   
\begin{equation}
\varrho^{(k)}_{C,1...k} = \frac{N!}{(N-k)!}P^{(k)}_{1...k}, 
\label{eq:003}
\end{equation}
where $\varrho^{(k)}_{C,1...k}$ sets the probability density to find a given configuration of $k$ arbitrary particles for the canonical ensemble.

\subsection{\label{subsec:02b}Grand canonical ensemble (GCE)}

Let us average equation (\ref{eq:003}) over the fluctuations of the number of particles, that is, let us apply operation $\sum_{N=0}^\infty P_N^V$ to both of its sides, where
\begin{equation}
P_N^V = \frac{z^N Z_N}{N!\Xi_V} 
\label{eq:004}
\end{equation}
is the probability for the GCE to have a definite number of particles $N$ inside volume $V$. Here $z$ is activity:
\begin{equation}
z = \frac{e^{\mu/k_BT}}{\Lambda^3}, 
\label{eq:005}
\end{equation}
where $\mu$ is the chemical potential, $\Lambda = h/\sqrt[]{2 \pi mk_BT}$, $h$ is Planck constant, $m$ is the mass of a particle, and $\Xi_V$ is the grand partition function of the system having volume $V$:
\begin{equation}
\Xi_V = 1 + \sum_{N=1}^\infty \frac{z^N Z_N}{N!}. 
\label{eq:006}
\end{equation}

We obtain:
\begin{equation}
\varrho^{(k)}_{G,1...k} = \sum_{N=k}^\infty \varrho^{(k)}_{C,1...k} P_N^V,
\label{eq:007}
\end{equation}
or
\begin{eqnarray}
\varrho^{(k)}_{G,1...k}  =&& \frac{z^k}{\Xi_V}  \Big \{ \exp(-\beta U^{k}_{1...k})  +  \sum_{N=1}^\infty \frac{z^N}{N!} \label{eq:008} \\
&&    \times  \int \limits_V   \exp(-\beta U^{N+k}_{1...N+k})d\bm{r}_{k+1}...d\bm{r}_{k+N} \Big \}  \nonumber.
\end{eqnarray}

$\varrho^{(k)}_{G,1...k}$ specifies the probability density to find  certain configuration of $k$ arbitrary particles for the GCE. For ideal gas $\varrho^{(k)}_{G,1...k} = \varrho^{k}$, where $\varrho = \overline{N}/V$ is the number density.

\subsection{\label{subsec:02c}Ursell factors and factors of partial localization}

Ursell factors ${\cal U}^{(k)}_{1...k}$ are also called cluster functions, it is these functions that appear in the known expansion of pressure in powers of the activity \cite[p.129]{hillstatmeh1987}, \cite[p.232]{landaulifshitz1985}
\begin{equation}
P(z,T) =zk_BT  + k_BT \sum_{k=2}^\infty \frac{z^k}{k!} \int {\cal U}^{(k)}_{1...k} d\bm{r}_2...d\bm{r}_k.
\label{eq:009} 
\end{equation}

They are also included into the corresponding expansion of the number density
\begin{equation}
\varrho(z) = z + z\sum_{n=1}^\infty \frac{z^n}{n!} \int {\cal U}^{(n+1)}_{1...n+1} d\bm{r}_{2}...d\bm{r}_{n+1},
\label{eq:010}
\end{equation}
which is evident from the relation \footnote[1]{To avoid bulky equations, we will not indicate constant temperature; this will be implied for all the derivatives. In addition, differentiation parameters will not be indicated in the cases when their values are evident, for example are defined by the opposite side of an equation.}
\begin{equation}
\varrho = \frac{\partial P}{\partial \mu}
\label{eq:011}.
\end{equation}

Ursell factors possess the locality property which is essential for this consideration: they decay rapidly with the removal of any group of particles including the unitary one. 

Taking into account that (\ref{eq:009}) is in fact the expansion in Taylor series, and keeping in mind the local character of Ursell factors we may write down the equation 
\begin{equation}
{\cal U}^{(k)}_{1...k}  = \beta \frac{\partial^{k}P}{\partial z^{k} } \prod_{n = 2}^k \delta(\bm{r}_n - \bm{r}_1),
\label{eq:012}
\end{equation}
which is true on the macroscopic scale and where the derivatives are taken at the point $z = 0$. Here $\delta(\bm{r})$ is the Dirac delta function. We naturally assume that index $k$ does not reach the macroscopic values.

Some other properties of Ursell factors are described in Appendix \ref{subsec:appenda1}.

The factors of partial localization ${\cal B}^{(m,k)}_{1...m+k}$ are the hybrids of Botlzmann and Ursell factors. They are included in the expansion of the distribution functions in powers of the activity
\begin{eqnarray}
\varrho^{(m)}_{G,1...m} (\psi^V)&& = z^m \Big \{ {\cal B}^{(m,0)}_{1...m}  +  \sum_{k=1}^\infty \frac{z^k}{k!} \label{eq:013} \\
 \times &&   \int   \Big [ \prod_{i = m+1}^{m+k} \psi^V_i \Big ]   {\cal B}^{(m,k)}_{1...m+k} d\bm{r}_{m+1}...d\bm{r}_{m+k} \Big \}  \nonumber,
\end{eqnarray}
where
\begin{equation}
\psi^{V}_i \equiv \psi^{V}(\bm{r}_i) = 
	 \left\{ 
			\begin{array}{ll} 
         1 & (\bm{r}_i \in V)\\   
         0 & (\bm{r}_i \notin V)
     	\end{array}  
		\right.
			\label{eq:014}
\end{equation}
are characteristic functions.

Some properties of ${\cal B}^{(m,k)}_{1...m+k}$ are described in Appendix \ref{subsec:appenda2}.

\subsection{\label{subsec:02d}Ursell functions}

Ursell functions are analogous to Ursell factors but they are composed not on the basis of Boltzmann factors but on the basis of distribution functions of the type (\ref{eq:008}) \cite{percus1964}. They are also called localized distribution functions.

A definition of Ursell functions is the equality similar to (\ref{eq:a01}) and true for arbitrary $k \geq 1$:
\begin{eqnarray}
{\cal F}^{(k)}_{1...k} &=& \sum_{\{\bm{n}\}}(-1)^{l-1}(l-1)!\prod_{\alpha = 1}^l \varrho^{(k_\alpha)}(\{\bm{n}_\alpha\}), \nonumber \\
1 & \leq&  k_\alpha \leq k, ~~~~~~~ \sum_{\alpha = 1}^l k_\alpha = k, \label{eq:015}
\end{eqnarray}
where $\{\bm{n}\}$ designates some partitioning of the given set of $k$ particles with coordinates $\bm{r}_1,...\bm{r}_k$ into disjoint groups $\{\bm{n}_\alpha\}$, $l$ is the number of groups of a specific  partitioning, $k_\alpha$ is the size of the group numbered $\alpha$, summing is performed over all possible partitionings. 

These functions also possess the property of locality: they decay rapidly with the removal of any group of particles including a single particle. Several first functions for the uniform medium look like 
\begin{eqnarray}
{\cal F}^{(1)}_{1}\mkern 5mu  &= & \varrho\label{eq:016}  \\
{\cal F}^{(2)}_{1,2} \mkern 6mu &= & \varrho^{(2)}_{1,2} -  \varrho^2 \nonumber \\
{\cal F}^{(3)}_{1,2,3}  &= & \varrho^{(3)}_{1,2,3} -  ( \varrho^{(2)}_{1,2} + \varrho^{(2)}_{2,3} + \varrho^{(2)}_{1,3})\varrho  + 2\varrho^{3} \nonumber \\
\dotso \nonumber, 
\end{eqnarray}
similarly to (\ref{eq:a02}). 

We will need the expansion for the uniform medium 
\begin{equation}
{\cal F}^{(k)}_{1...k}(z)= z^k \Big \{ {\cal  U}^{(k)}_{1...k} + \sum_{n=1}^\infty \frac{z^n}{n!}\int  {\cal  U}^{(n+k)}_{1...n+k} d\bm{r}_{k+1}...d\bm{r}_{k+n} \Big \}, 
\label{eq:017}
\end{equation}
connecting them with Ursell factors, as obtained in \cite{UhlenbeckFord1962}. It should be stressed that integration is carried out in (\ref{eq:017}) over the whole space \cite{zaskulnikov200911a}. Note that another kind of Ursell factors, an equivalent one, was also used in \cite{UhlenbeckFord1962}.

\subsection{\label{subsec:02e}Presence of an external field}

The configuration integral of a nonuniform closed system is given by expression 
\begin{equation}
Z^U_N = \int\limits_{V}^{} \exp(-\beta\sum_{i=1}^N v_i-\beta U^N_{1...N}) d\bm{r}_1...d\bm{r}_N, \\
\label{eq:018}
\end{equation}
where $v_i$ is the energy ($=$ potential) of the interaction of the $i$-th particle with the field.    

For the GCE, we introduce the quantity
\begin{equation}
\Xi^U_V = 1+ \sum_{N=1}^\infty \frac{z^N Z^U_N}{N!}, 
\label{eq:019}
\end{equation}
which is evidently the grand partition function of the system in the presence of an external field.

Using the properties of the fractional generating function (\ref{eq:008}) (see Appendix \ref{subsec:appenda2}, \ref{subsec:appendb1}) we obtain an analog of (\ref{eq:010})
\begin{equation}
\varrho(\bm{r_1},z) = z \theta_1 + z\theta_1\sum_{n=1}^\infty \frac{z^n}{n!} \int \Big [ \prod_{i = 2}^{n+1} \theta_i \Big ]{\cal U}^{(n+1)}_{1...n+1} d\bm{r}_{2}...d\bm{r}_{n+1},
\label{eq:020}
\end{equation}
where $\varrho(\bm{r_1},z)$ is the number density in the presence of an external field, and
\begin{equation}
\theta_i =  \exp(-\beta v_i) = \exp[-\beta v(\bm{r_i})],
\label{eq:021}  
\end{equation}
and the analog of (\ref{eq:013})
\begin{eqnarray}
\varrho^{(m)}_{1...m} (\theta)&& = z^m \Big  [ \prod_{i = 1}^{m} \theta_i \Big  ] \Big  \{ {\cal B}^{(m,0)}_{1...m}  +  \sum_{k=1}^\infty \frac{z^k}{k!}  \label{eq:022} \\
&&   \times  \int  \Big  [ \prod_{i = m+1}^{m+k} \theta_i \Big  ]   {\cal B}^{(m,k)}_{1...m+k} d\bm{r}_{m+1}...d\bm{r}_{m+k} \Big \}.  \nonumber
\end{eqnarray}

\section{\label{subsec:03a}Basic equations}

We will consider an open statistical system in the field, with the scope of the field smaller than the size of the system itself, and the position of the field at rather large distance from the boundaries of the system, that is, we talk about a situation more likely denoted as ``the field in the system''.

We will assume that the potential of the field is constant, except for the boundary regions, unless otherwise stated. Its changes on the macroscopic scale can be considered admissible.

The potential of external fields will be considered to increase/decay sufficiently rapidly to provide the convergence of the corresponding integrals when moving into the depth of the body or outside, while the interparticle potential decays sufficiently rapidly to ensure the convergence of the zero and first moments of Ursell factors (so, we are not considering Coulomb potentials in this work).

In some cases we will accept that the region affected by the field of the solid body has a finite radius, and speak of the transition region between the solid body and the fluid. This approximation in most cases does not have a principal character and will be used to simplify consideration. 

Finally, unless otherwise stated, we will accept that the region of the transition from the solid body to the fluid is much smaller than the linear size of the field, and the profile of the field along its gradient is the same everywhere. 

Some sections of this paper are similar to the corresponding sections of \cite{Zaskulnikov201102a}. 

Taking the logarithm of the ratio of the partition functions (\ref{eq:019}) and (\ref{eq:006}), we obtain
\begin{equation}
\mkern -200mu \ln{\Xi^U_V}-\ln{\Xi_V} = \sum_{k=1}^\infty \frac{z^k}{k!}
\label{eq:023}
\end{equation}
\[
\times \int\limits_{V}^{}\Big [\exp(-\beta\sum_{i=1}^k v_i)-1 \Big ] {\cal U}^{(k)}_{1...k} d\bm{r}_1...d\bm{r}_k,
\]
where we have used a logarithmic type of the generating function of the Ursell factors.

In the above-indicated assumptions, we may remove symbol $V$ from the integrals over the coordinates of particles in (\ref{eq:023}) and accept that integration is carried out over the infinite space. 

Equation (\ref{eq:023}) is transformed as follows:    

\begin{widetext}
\begin{equation}
\ln{\Xi^U_V}-\ln{\Xi_V} = - \sum_{k=1}^\infty \frac{z^k - z'^k}{k!} \int \widetilde{\varphi}_1 {\cal U}^{(k)}_{1...k} d\bm{r}_1...d\bm{r}_k+ \sum_{k=2}^\infty \frac{1}{k!} \int \Big [ \prod_{i = 1}^k (z \widetilde{\theta}_i + z' \widetilde{\varphi}_i) 
 - \widetilde{\theta}_1 z^k - \widetilde{\varphi}_1 z'^k \Big ]{\cal U}^{(k)}_{1...k} d\bm{r}_1...d\bm{r}_k,
\label{eq:024}
\end{equation}
\end{widetext}
where
\begin{equation}
z' = z \exp {(- \beta v_0)}
\label{eq:025}
\end{equation}
is activity, and $v_0$ is the potential of the external field inside the region affected by the field at the sufficient remoteness from the boundaries, and where designations  
\begin{equation}
\widetilde{\theta}_i = \frac{\exp {(- \beta v_i)} - \exp {(- \beta v_0)}}{1 - \exp {(- \beta v_0)}}
\label{eq:026}
\end{equation}
and
\begin{equation}
\widetilde{\varphi}_i = \frac{1 - \exp {(- \beta v_i)} }{1 - \exp {(- \beta v_0)}}, 
\label{eq:027}
\end{equation}
are introduced, generalizing Boltzmann factors (\ref{eq:021}) and Mayer functions
\begin{equation}
\varphi_i = 1 - \exp {(- \beta v_i)} 
\label{eq:028}
\end{equation} 
respectively. One can see that 
\begin{equation}
\widetilde{\theta}_i + \widetilde{\varphi}_i = 1
\label{eq:029}
\end{equation}
and functions $\widetilde{\theta}_i$ and $\widetilde{\varphi}_i$ decay inside and outside the field, respectively, and tend to 1 in the contrary case. 

In addition, they have the property 
\begin{equation}
z\widetilde{\theta}_i + z'\widetilde{\varphi}_i = z\theta_i
\label{eq:030}.
\end{equation}

(Symmetry property of these functions (\ref{eq:120}), see section \ref{subsec:03l}.)

The first sum in right-hand side of (\ref{eq:024}), as one can see from (\ref{eq:009}), is proportional to the difference of pressures of uniform systems. 

The second term in the right-hand side of (\ref{eq:024}) is proportional to the area of the surface separating the field region from the rest part of the system. Indeed, its structure is such that at least two particles always stay on different sides of the ``field boundary''. If we expand the product in the arbitrary term in (\ref{eq:024}), we will obtain the expressions of the type of 
\begin{equation}
\int \Big [\prod_{\alpha = 1}^{k} \widetilde{\theta}_{i_\alpha} \prod_{\beta = 1}^{l}  \widetilde{\varphi}_{j_\beta} \Big ] {\cal U}^{(t)}_{1...t} d\bm{r}_1...d\bm{r}_t,
\label{eq:031}  
\end{equation}
where $1 \leq k \leq t -1 ; ~ 1 \leq l \leq t - 1; ~ k+l=t$. Assume that the first particle belongs to group $\widetilde{\theta}_i$.  (This may be made without loss of generality due to the symmetry of Ursell factors under permutations). With the fixed first particle, due to the locality of Ursell factors, integrals of the type of (\ref{eq:031}) are defined by the local area near it. As a consequence, $\int \dots d\bm{r}_2\dots d\bm{r}_t$ are independent of the displacement of the first particle along the boundary of the system. 

With the displacement of the first particle from the boundary to the depth of the field region, they decay rapidly due to factor $\widetilde{\theta}_1$. With the displacement of the first particle from the boundary to the depth of the free fluid, they decay rapidly due to fixing factors $\widetilde{\varphi}_j$ and the local nature of Ursell factors. Integrating along the surface, we can factor out of the sum the area and make other simple transformations to obtain 
\begin{equation}
\Omega^U = -P(z)V +[P(z)-P(z')]\int \widetilde{\varphi}(\bm{r}) d\bm{r} + \nu(z, z') A,
\label{eq:032}
\end{equation}
where $\Omega^U$ is the omega- potential of the system with the field immersed into it, $A$ is the area limiting the field region and where we introduced designation 
\begin{eqnarray}
&&\nu (z,z') =  k_BT \sum_{k=2}^\infty \frac{1}{k!}
\label{eq:033} \\ 
&&\times \int\Big [ \widetilde{\theta}_1 z^k + \widetilde{\varphi}_1 z'^k -\prod_{i = 1}^k (z \widetilde{\theta}_i + z' \widetilde{\varphi}_i)\Big ]{\cal U}^{(k)}_{1...k} dx_1 d\bm{r}_2...d\bm{r}_k \nonumber.
\end{eqnarray}

Here $x_1$ is the coordinate directed along the field gradient, and the direction of $x_1$ axis is chosen so that $dx_1 > 0$.

To remind, integration is carried out over the infinite space. 

In (\ref{eq:033}), integration over $x_1$ may be performed. Changing the variables $x'_1 = x_1,~\bm{r}'_i = \bm{r}_i - \bm{r}_1,~ i=2,\dots k$ and using the fact that the Jakobian of this transformation is equal to $1$ and the invariance of ${\cal U}^{(k)}_{1...k}$ under translations, we obtain 
\begin{equation}
\nu (z,z') =  k_BT \sum_{k=2}^\infty \frac{1}{k!}\int {g}^{(k-1)}_{2...k}  {\cal U}^{(k)}_{0,2...k} d\bm{r}_2...d\bm{r}_k,
\label{eq:034}
\end{equation}
where
\begin{eqnarray}
&&{g}^{(k-1)}_{2...k}= \int \limits_{- \infty}^{+ \infty}   \Big \{\widetilde{\theta}(x) z^k +\widetilde{\varphi}(x) z'^k   \label{eq:035}  \\
&&~~  -[z \widetilde{\theta}(x) + z' \widetilde{\varphi}(x)]\prod_{i = 2}^k [z \widetilde{\theta}(x+x_i) + z' \widetilde{\varphi}(x+x_i)] \Big \} d x
\nonumber
\end{eqnarray}
is the function of variables $x_2,...x_k$, symmetric under permutations. Evidently, unlike ${\cal U}^{(k)}_{1...k}$ it is not invariant under translations because it depends on the localization of the external field. 

It should be noted that the variables in (\ref{eq:034}) are separated: the external potential is present only in ${g}^{(k-1)}$, while the interparticle potential is present only in ${\cal U}^{(k)}$. 

We will see below (section \ref{subsec:03l}) that coefficient $\nu$ is symmetrical under permutation of $z$ and $z'$, with the corresponding changes of the level of energy readout and its sign.

Equations (\ref{eq:032}) and (\ref{eq:034}) generalize expressions for the $\Omega$- potential and nonlinear surface coefficient \cite{Zaskulnikov201102a} to the case of the finite field. They are basic for subsequent consideration. The first term in the right side of (\ref{eq:032}) has purely volume properties, and the third one has purely surface properties.  The second term as is easily seen from its structure possesses both volume and surface properties. 

As can be clearly seen, the expansion of a function $\nu (z,z')$ in series of $z, z'$ begins with the quadratic form. Other representations of this quantity will be established below.

\section{\label{subsec:03b}Variants of accounting for surface effects}

Equation (\ref{eq:032}) may be transformed to the form:
\begin{eqnarray}
\Omega = -P(z)(V-V') - P(z')V' &&  \nonumber  \\
- \big [P(z)-P(z') \big ]\Big [\int\limits_{-\infty}^{x'}\widetilde{\theta}(x) dx &-& \int\limits_{x'}^{\infty}\widetilde{\varphi}(x) dx\Big ] A \nonumber \\
&&+ \nu(z,z') A,\label{eq:036}
\end{eqnarray} 
where $x'$ is some arbitrary point inside the transition layer or near it, determining the volume $V'$, and $x$ is the coordinate directed along the field gradient. (Equally well, the regions of integration in equation (\ref{eq:036}) can be bounded by the limits of the transition layer because the integrands become zero outside it.)

We see that there is a definite degree of arbitrariness in partitioning the volume and surface terms. In reality, this freedom exists only within the transition layer or near it. Indeed, one can easily understand that otherwise we speak of the trivial compensation of two identical terms having opposite signs. 

Let us denote the volume of the constant field as $V_1$, and the volume of the transition region as $V_t$. As we have already mentioned, $V_t \ll V_1$.

At the first glance, from (\ref{eq:036}) we obtain three versions of accounting for surface effects for different  values $x'$:
\begin{eqnarray}
\Omega = -&&P(z)(V-V_1-V_t) - P(z')(V_1+V_t)   \nonumber  \\
&&- [P(z)-P(z')]A\int\limits_{l_t}^{}\widetilde{\theta}(x) dx + \nu(z,z') A,\label{eq:037} 
\end{eqnarray} 
\begin{eqnarray}
\Omega = -&&P(z)(V-V_1) - P(z')V_1   \nonumber  \\
&&+ [P(z)-P(z')]A\int\limits_{l_t}^{}\widetilde{\varphi}(x) dx + \nu(z,z') A,\label{eq:038} 
\end{eqnarray}
\begin{equation}
\Omega = -P(z)(V-V_0) - P(z')V_0    + \nu(z,z') A,
\label{eq:039}
\end{equation}
where $l_t$ is the length of the transition region, and $V_0$ is the volume of the body limited by the surface that is set's by condition 
\begin{equation}
\int\limits_{-\infty}^{x_0} \widetilde{\theta}(x) dx = \int\limits_{x_0}^{\infty} \widetilde{\varphi}(x) dx. 
\label{eq:040}
\end{equation}
or
\begin{equation}
\int\limits_{-\infty}^{x_0}[ \exp {(- \beta v)} - \exp {(- \beta v_0)}] dx = \int\limits_{x_0}^{\infty} [1- \exp {(- \beta v)} ] dx. 
\label{eq:041}
\end{equation}

In the left side of this equality, integration is carried out from the field region outwards, while in the right part it is made to the depth of the fluid, and it determines the position of the surface of interest for us $x_0$. 

The variant (\ref{eq:037}) corresponds to $x'$ being at the boundary of the uniform fluid, that is, relates the surface effects to the solid body (field), while (\ref{eq:038}) - to $x'$ at the boundary of the uniform solid body, and relates the surface effects to the liquid. Variant (\ref{eq:039}) makes the linear surface terms equal to zero using the property (\ref{eq:040}). Depending on the shape of potential, $x'$ may be either inside the transition region or outside it. 

In reality, due to the symmetry of the problem of permeable wall with respect to the permutation of the fluid and field (see section \ref{subsec:03l}), we have only two different versions. The variants (\ref{eq:037}) and (\ref{eq:038}) are equivalent in this case. This can be easily proved by making substitutions in (\ref{eq:037}) as $V_1^* = V - V_1 - V_t$  and (\ref{eq:119}) - (\ref{eq:121}).

Of course, other version based on equation (\ref{eq:036}) can be considered. The results obtained here we will discuss below.

Expression (\ref{eq:036}) is equivalent to
\begin{equation}
\Omega = -P(z)(V-V') - P(z')V'  + \gamma  A,
\label{eq:042}
\end{equation}
where
\begin{equation}
\gamma = - [P(z)-P(z')]\Big [\int\limits_{-\infty}^{x'}\widetilde{\theta}(x) dx - \int\limits_{x'}^{\infty}\widetilde{\varphi}(x) dx\Big ] + \nu (z,z') 
\label{eq:043}
\end{equation}
is the general form of the specific surface $\Omega$-potential. This expression is separated into the linear and non-linear ($\nu$) terms with respect to pressure.

\section{\label{subsec:03c}Surface number density}

Differentiating (\ref{eq:036}) with respect to the chemical potential and taking into account the fact that \footnote[2]{We will omit the averaging sign for $N$ of open systems.}
\begin{equation}
N = - \bigg (  \frac{\partial \Omega}{\partial \mu} \bigg )_{V,A},   
\label{eq:044} 
\end{equation} 
and keeping in mind (\ref{eq:011}), we obtain 
\begin{eqnarray}
N = \varrho(z)(V-V') + \varrho(z')V' &&  \nonumber  \\
+ \big[\varrho(z)-\varrho(z') \big ]\Big [\int\limits_{-\infty}^{x'}\widetilde{\theta}(x) dx &-& \int\limits_{x'}^{\infty}\widetilde{\varphi}(x) dx\Big ] A \nonumber \\
&&- A\beta z \frac{\partial\nu(z,z')}{\partial z }.\label{eq:045}
\end{eqnarray} 

Differentiating $\nu(z,z')$ with respect to $z$ it is evidently necessary to consider $z'$ as a function of $z$ in accordance with (\ref{eq:025}).

Expression (\ref{eq:045}) can be written in the form 
\begin{equation}
N = N_{b_1} + N_{b_2} + N_s,
\label{eq:046}  
\end{equation}
where $N_{b_1}, N_{b_2}$ and $N_s$ are the numbers of volume particles in the different phases and the surface particles, respectively. For them, we have the expressions
\begin{equation}
N_{b_1} =  \varrho(z)(V-V'), ~ N_{b_2} = \varrho(z')V',
\label{eq:047}  
\end{equation}
\begin{eqnarray}
N_s= \big [\varrho(z)-\varrho(z')\big ]\Big [\int\limits_{-\infty}^{x'}\widetilde{\theta}(x) dx &-& \int\limits_{x'}^{\infty}\widetilde{\varphi}(x) dx\Big ] A \nonumber \\
&-& A\beta z \frac{\partial\nu(z,z')}{\partial z }.\label{eq:048}
\end{eqnarray} 

Passing to the surface number density
\begin{equation}
\varrho_s =   \frac{N_s}{A},
\label{eq:049}  
\end{equation}
we obtain 
\begin{equation}
\varrho_s = \big [\varrho(z)-\varrho(z')\big ]\Big [\int\limits_{-\infty}^{x'}\widetilde{\theta}(x) dx - \int\limits_{x'}^{\infty}\widetilde{\varphi}(x) dx\Big ] - \beta z \frac{\partial\nu(z,z')}{\partial z }.
\label{eq:050}
\end{equation}

Taking into account that from (\ref{eq:020}) follows that
\begin{equation}
- \beta z \frac{\partial\nu(z,z')}{\partial z} =   \int\limits_{-\infty}^{\infty}\left [ \varrho(\bm{r}) - \widetilde{\theta}(x) \varrho(z) - \widetilde{\varphi}(x) \varrho(z') \right ] dx,
\label{eq:051}  
\end{equation}
we obtain finally after simple transformations
\begin{equation}
\varrho_s =   \int\limits_{-\infty}^{x'} \left [ \varrho(x) - \varrho(z') \right ] dx + \int\limits_{x'}^{\infty}\left [ \varrho(x) - \varrho(z) \right ] dx,
\label{eq:052}  
\end{equation}
where $\varrho(x)$ is the number density in the vicinity of the field boundaries depending on the coordinate directed along the field gradient.

We will see further on (section \ref{subsec:03g}) that (\ref{eq:052}) agrees with the usual expression 
\begin{equation}
\varrho_s = - \frac{\partial \gamma}{\partial \mu}.
\label{eq:053}
\end{equation}

Formula (\ref{eq:052}) generalizes the expression for the surface number density \cite{Zaskulnikov201102a} to the case of the field of finite value.

Below we obtain the second form for $\varrho_s$ (\ref{eq:072}), in which the terms are grouped according to linearity criterion.

Expression (\ref{eq:052}) has a simple sense. Relying on the evident relation 
\begin{equation}
N =   \int\limits_{V}^{} \varrho(\bm{r}) d\bm{r}
\label{eq:054}  
\end{equation}
and using (\ref{eq:046}) and (\ref{eq:047}), we obtain 
\begin{eqnarray}
N_s = && N - N_{b_1} - N_{b_2}  \nonumber \\
&& =\int\limits_{V}^{} \varrho(\bm{r}) d\bm{r} - \varrho(z)(V-V')  - \varrho(z')V'\label{eq:055} \\
&& = \int\limits_{V'}^{} \left [ \varrho(\bm{r}) - \varrho(z') \right ] d\bm{r} + \int\limits_{V-V'}^{} \left [ \varrho(\bm{r}) - \varrho(z) \right ] d\bm{r},
\nonumber  
\end{eqnarray}
which under the above assumptions is identical to (\ref{eq:052}). So, this expression is true for any densities of the fluid. The correspondence between (\ref{eq:052}) and (\ref{eq:055}) provides evidence of the internal consistency of the approach applied. 

To conclude, it should be noted that, unlike in the case of the equilibrium system liquid/gas, similarly to the case of impermeable wall \cite{Zaskulnikov201102a}, superimposition of the condition $N_s = 0$, as one can see from (\ref{eq:052}), results in the dependence of the position of partitioning surface on number density. So, in this case, too, the physical sense of such a partition is absent. Instead, in some cases it is possible to use the condition (\ref{eq:040}), coming to the variant of ``zero'' adsorption.

\section{\label{subsec:03d}Henry constants of adsorption and absorption}

For Henry constant of adsorption
\begin{equation}
K_H = \lim_{\varrho\rightarrow 0} \frac{\varrho_s}{\varrho(z)} 
\label{eq:056}  
\end{equation} 
from (\ref{eq:050}) or (\ref{eq:052}) we obtain
\begin{eqnarray}
 K_H =  \int\limits_{-\infty}^{x'} \big [ \exp(-\beta v)&& - \exp(-\beta v_0) \big ] dx  \nonumber \\
&& - \int\limits_{x'}^{\infty} \big [ 1 - \exp(-\beta v) \big ] dx.
\label{eq:057}
\end{eqnarray}

This formula generalizes the previously obtained expression \cite{Zaskulnikov201102a} to the case of the field of finite value.

The dependence of $K_H$ on $x'$ is universal and can always be extracted in the explicit form. Indeed, subtracting one value of the constant from another we make sure that
\begin{equation}
K_H(x') = (x' - x'')[1 - \exp(-\beta v_0)]+ K_H(x'').
\label{eq:058}
\end{equation}

If we choose the value $x'' = x_0$, determined by (\ref{eq:040}) or (\ref{eq:041}), then, because
\begin{equation}
K_H(x_0) = 0
\label{eq:059}
\end{equation} 
we obtain
\begin{equation}
K_H(x') = [x' - x_0(T)][1 - \exp(-\beta v_0)]
\label{eq:060}
\end{equation}
and the problem of $K_H$ determination is reduced to the calculation of $x_0$. 

The sense of expression (\ref{eq:060}) is quite clear: this is a coefficient proportional to the difference of densities in the limit of low activities: $[1 - \exp(-\beta v_0)]$, multiplied by the volume of the parallelepiped with thickness $[x' - x_0(T)]$ and unit base area.

It follows from (\ref{eq:040}) that 
\begin{equation}
x_0 = - \int\limits_{-\infty}^{0} \widetilde{\theta}(x) dx + \int\limits_{0}^{\infty} \widetilde{\varphi}(x) dx. 
\label{eq:061}
\end{equation}

The surface passing through point $x_0$ by virtue of (\ref{eq:059}) can be called the surface of zero adsorption. 

For Henry constant of absorption, using expansion (\ref{eq:020}), we obtain
\begin{equation}
k_H = \lim_{\varrho\rightarrow 0} \frac{\varrho(z')}{\varrho(z)} = \exp(-\beta v_0)
\label{eq:062}.
\end{equation}

Equations (\ref{eq:057}), (\ref{eq:062}) determine, within the framework of the approach applied, the connection between the two Henry constants: adsorption and absorption. Here we speak of the parametric dependence on temperature under the fixed potential.

Indeed, excluding temperature from equations (\ref{eq:057}), (\ref{eq:062}) we obtain 
\begin{equation}
K_H =  \int\limits_{-\infty}^{x'}\big [ k_H^{ \widetilde{v}(x)} - k_H\big ] dx - \int\limits_{x'}^{\infty}\big [ 1 - k_H^{\widetilde{v}(x)}\big ] dx,
\label{eq:063}  
\end{equation}
where
\begin{equation}
\widetilde{v}(x) = \frac{v(x)}{v_0}
\label{eq:064}
\end{equation}
is the potential of particle interaction with the external field, normalized so that it is equal to $1$ inside the body.

The procedure used here may seem somewhat artificial, but the key factor is that the reduced potential $\widetilde{v}$ in fact does not contain parameter $v_0$ as a consequence of binding the actual potential at the boundary of the body to the $v_0$ value. In other words, parameters $v_0$ and $\beta$ are included in expressions only in the combination $\beta v_0$ and thus the procedure is legal. 

Now we can make some general conclusions concerning the mutual behavior of $K_H$ and $k_H$. The relations that we have in mind are of the form
\begin{eqnarray}
k_H \rightarrow 1 &\Rightarrow & K_H \rightarrow 0  \label{eq:065} \\
k_H \rightarrow 0 &\Rightarrow & K_H \rightarrow x' - x_2 \label{eq:066}.
\end{eqnarray}

As one can easily make sure, these relations follow directly from (\ref{eq:063}).

The relationship (\ref{eq:065}) corresponds to high temperatures and agrees with the simple fact that the surface effects disappear in the uniform medium.

The relation (\ref{eq:066}) is true, provided that $v(x) > 0$ in the whole space. This means also that $v_0 > 0$. The value $x_2$ corresponds to the point in which the external field becomes zero. So, this expression is true only under the assumption of the finite thickness of the transition layer. (Due to the symmetry with respect to the substitution field $\leftrightarrow$ fluid (section \ref{subsec:03l}) the case under consideration in fact includes also the case $v(x) < 0$. That is, here we exclude only the variant of sign change for $v(x)$.)

The expression (\ref{eq:066}) has a simple sense: for $x' > x_2$ this is the volume of a parallelepiped that has the base of the unit area and the height $x' - x_2$. Multiplying it by the volume number density we obtain the surface number density. (Of course, we consider the limit of low densities.) The sense of $K_H$ for $x' < x_2$ is clear too: this is the negative amount compensating the fictive amount of fluid in the region of the solid body.

So, provided temperature tends to zero in the point where the field vanishes ($x_2$), in fact, an infinite wall appears.

It is clear from the above considerations that the equations (\ref{eq:066}) give the maximal value of constant $K_H$ for non-negative potentials under $x' > x_2$.

Thus, expressions (\ref{eq:065}) and (\ref{eq:066}) show that adsorption and absorption anticorrelate in the case of the absence of the regions of negative potential.

In (\ref{eq:063}), terms with $x'$ may be distinguished; then 
\begin{equation}
K_H = x'(1 - k_H) + \int\limits_{-\infty}^{0}\big [ k_H^{\widetilde{v}(x)} - k_H\big ] dx - \int\limits_{0}^{\infty}\big [ 1 - k_H^{\widetilde{v}(x)}\big ] dx
\label{eq:067}.  
\end{equation}

As an example, consider the case of the linear change of potential in the transition region 
\begin{equation}
\widetilde{v}(x) = 
	 \left\{ 
			\begin{array}{ll} 
         1 & (x < x_1)\\   
         (x_2 - x)/(x_2 - x_1) & (x_1 < x < x_2) \\
         0 & (x_2 < x)
     	\end{array}  
		\right.
		\label{eq:068},
\end{equation}
where $x_1$ and $x_2$ are the coordinates of the beginning and end of the transition region from the body to the fluid.

Carrying out an elementary integration we obtain 
\begin{equation}
K_H = x'(1 - k_H) - \frac{1 - k_H}{\ln k_H} (x_2 - x_1) + x_1 k_H - x_2
\label{eq:069}.  
\end{equation}

It should be stressed once more that the above consideration relates to the parametric dependence of $K_H$ on $k_H$. For the direct dependence through $v_0$, the situation is more complicated because it is not clear how to model the behavior of the potential $v(x)$ in the transition region under variations of $v_0$. From the viewpoint of expression (\ref{eq:060}), the difficulty is that $x_0$ depends on $v_0$ in a non-trivial manner. In all likelihood, this dependence can be considered only within the frameworks of specific models the significance of which is unobvious.

\section{\label{subsec:03e}The basic form}

It follows from (\ref{eq:036}) that
\begin{eqnarray}
\Omega =&& -P(z)(V-V') - P(z')V'   \nonumber  \\
&&- \big[ P(z)-P(z')\big ](x'-x_0) A + \nu(z,z') A,
\label{eq:070}
\end{eqnarray} 
or, by virtue of  (\ref{eq:042})
\begin{equation}
\gamma = - \big [P(z)-P(z')\big ](x'-x_0) + \nu (z,z') 
\label{eq:071}.
\end{equation}

This equality is one of the basic expressions for the consideration of surface phenomena in this case. It binds four main surface parameters: specific surface  $\Omega$-potential $\gamma$, the position of the planes of zero adsorption $x_0$ and separating surface $x'$,  nonlinear surface coefficient $\nu$ and volume parameters - pressure at both sides of the interface. This equality also generalizes the analogous equality \cite{Zaskulnikov201102a} to the case of the field of finite value.

Differentiating (\ref{eq:071}) with respect to the chemical potential and taking into account (\ref{eq:051}) we obtain 
\begin{eqnarray}
 \varrho_s = && \big [\varrho(z) - \varrho(z')\big ] (x'-x_0) \nonumber \\
&& + \int\limits_{-\infty}^{+\infty}\big [ \varrho(x) - \widetilde{\theta}(x) \varrho(z) - \widetilde{\varphi}(x) \varrho(z') \big ]  dx
\label{eq:072}
\end{eqnarray}
- one more expression for the surface number density in the case of the arbitrary position of the dividing surface. Here two parts are clearly separated: one of them is linear with respect to density and the other is nonlinear. (This statement is true if we consider $z' = z'(z)$, and the external potential is fixed. If we accept $z$ and $z'$ as independent variables, then the first term of the right part of (\ref{eq:072}) is evidently a linear form with respect to density, while the second term starts with the quadratic form of $\varrho(z), \varrho(z')$, which follows from (\ref{eq:051}) and (\ref{eq:034}), (\ref{eq:035}).) One can easily make sure that (\ref{eq:072}) coincides with (\ref{eq:052}). 

So, we may consider the separation of surface values according to the linearity principle on the same ground as that for the localization principle. Indeed, if we define the linear surface number density as
\begin{equation}
\varrho_{s,l} = \big [\varrho(z) - \varrho(z') \big ] (x'-x_0),
\label{eq:073}
\end{equation}
and the nonlinear one as
\begin{equation}
\varrho_{s,n} = \int\limits_{-\infty}^{+\infty} \big [ \varrho(x) - \widetilde{\theta}(x) \varrho(z) - \widetilde{\varphi}(x) \varrho(z') \big ]  dx,
\label{eq:074}
\end{equation}
then, taking into account  (\ref{eq:011}) and (\ref{eq:051}), we obtain the analogs of (\ref{eq:053})
\begin{equation}
\varrho_{s,l} = \frac{\partial \Big \{ \big [P(z)-P(z')\big ](x'-x_0)\Big \}}{\partial \mu} 
\label{eq:075}
\end{equation}
and
\begin{equation}
\varrho_{s,n} = - \frac{\partial \nu(z,z')}{\partial \mu},
\label{eq:076}
\end{equation}
according to (\ref{eq:053}), (\ref{eq:071}). To remind, in (\ref{eq:075}), (\ref{eq:076}) we differentiate both $z$, and $z'$. Naturally, 
\begin{equation}
\varrho_s = \varrho_{s,l} + \varrho_{s,n}
\label{eq:077}
\end{equation}
and expressions (\ref{eq:073}) - (\ref{eq:077}) determine the surface number density in the most general form.

\section{\label{subsec:03f}Macroscopically smooth field}

Let the characteristic distance at which the field undergoes essential changes be macroscopic. Substituting (\ref{eq:012}) into (\ref{eq:034}) and integrating over $\bm{r}_2, \dots,  \bm{r}_k$, we obtain 
\begin{eqnarray}
&& \nu (z,T) =   \sum_{k=2}^\infty \frac{z^k}{k!} \frac{\partial^{k}P}{\partial z^{k} }  \nonumber \\
&& \times\int \limits_{-\infty}^{\infty} \Big \{ \widetilde{\theta}(x)+ \widetilde{\varphi}(x)\exp(- k \beta v_0) -  \big [\theta(x)\big ]^k \Big \} dx 
\label{eq:078}
\end{eqnarray}
or
\begin{equation}
\nu (z,z') =  \int\limits_{-\infty}^{\infty}\left [ \widetilde{\theta}(x) P(z)+\widetilde{\varphi}(x)P(z')- P(\theta z)  \right ] dx.
\label{eq:079}
\end{equation}

Substituting this expression into (\ref{eq:032}), similarly to the case of infinite field \cite{Zaskulnikov201102a}, we arrive at a logical result 
\begin{eqnarray}
  \Omega^U =&&  -\int\limits_{V}^{} P(\theta z) d\bm{r} = -\int\limits_{V}^{} P(z e^{- \beta v}) d\bm{r} \nonumber \\
&& = -\int\limits_{V}^{} P[\mu - v(\bm{r})] d\bm{r} = -\int\limits_{V}^{} P(\bm{r}) d\bm{r},
\label{eq:080}
\end{eqnarray}
where we used a known expression for the chemical potential in the force field. 

We obtain from (\ref{eq:061}), (\ref{eq:071}) and (\ref{eq:079}) 
\begin{equation}
\gamma(x') =   \int\limits_{-\infty}^{x'}\big[P(z') - P(x)\big ] dx +  \int\limits_{x'}^{+\infty} \big [P(z) - P(x)\big ] dx.
\label{eq:081}
\end{equation}

A characteristic structure that appeared in (\ref{eq:081}), as will be seen below, is universal for $\gamma$.

\section{\label{subsec:03g}\texorpdfstring{``Local pressure''}{"Local pressure"}}%

Integrating (\ref{eq:044}) over the chemical potential we obtain
\begin{equation}
\Omega =  - \int\limits_{-\infty}^{\mu} N d\mu',
\label{eq:082}  
\end{equation}
where we omitted an arbitrary function of volume, temperature and field appearing as a result of integration because the  $\Omega$-potential should tend to zero in the low density limit. Using expression (\ref{eq:054}), we come to the relation 
\begin{equation}
\Omega =  - \int\limits_{V}^{} P^{*}(\bm{r}) d\bm{r},
\label{eq:083}  
\end{equation}
where
\begin{equation}
P^{*}(\bm{r}) =  \int\limits_{-\infty}^{\mu} \varrho(\bm{r}) d\mu',
\label{eq:084}  
\end{equation}
which is equivalent in this case to 
\begin{equation}
\varrho(\bm{r}) = \frac{\partial P^{*}(\bm{r})}{\partial \mu}.
\label{eq:085}
\end{equation}

This means that the number density for the system in the field and $P^{*}$ are connected through exactly the same relation as usual density and pressure for the uniform system (\ref{eq:011}). 

The value equivalent to $P^*$ was considered for the impermeable wall \cite{StillingerBuff1962}, \cite{Zaskulnikov201102a} and we will also call it ``local pressure''. 

Under the condition of Mayer- type cluster expansion validity, we obtain from (\ref{eq:020})
\begin{eqnarray}
P^{*}(\bm{r_1},z) &=&k_B T z \theta_1 \label{eq:086} \\
&+& k_B T \theta_1\sum_{n=2}^\infty \frac{z^n}{n!} \int \Big [ \prod_{i = 2}^{n} \theta_i \Big ]{\cal U}^{(n)}_{1...n} d\bm{r}_{2}...d\bm{r}_{n}, \nonumber
\end{eqnarray}
which also resembles the standard expansion of pressure in powers of the activity (\ref{eq:009}). 

One can see from (\ref{eq:086}) that $P^*$ rapidly tends to $P$ as the distance from the field sources increases. 

It should be stressed that equation (\ref{eq:083}) is absolutely general and true for the system in the force field of the arbitrary configuration. In the case of macroscopically smooth fields it turns into expression (\ref{eq:080}).

Similarly to the above consideration for the density of particles, equation (\ref{eq:083}) can be brought by means of simple transformations to the form 
\begin{eqnarray}
\Omega &=&   -P(z)(V - V') -P(z')V'\label{eq:087} \\
&& + \int\limits_{V'}^{}\big [P(z') - P^{*}(\bm{r})\big]d\bm{r} + \int\limits_{V - V'}^{} \big [P(z) - P^{*}(\bm{r})\big] d\bm{r}.
\nonumber
\end{eqnarray}

Integrating over the surface we obtain 
\begin{eqnarray}
\Omega &=&   -P(z)(V - V') -P(z')V'\label{eq:088} \\
&& + A \int\limits_{-\infty}^{x'} \big[P(z') - P^{*}(x)\big] dx + A \int\limits_{x'}^{\infty}\big [P(z) - P^{*}(x)\big] dx.
\nonumber
\end{eqnarray}

As it should be, the derivative of (\ref{eq:088}) with respect to the chemical potential gives (up to a multipliers) the expression for the volume number density (\ref{eq:047}) and the surface one (\ref{eq:052}).

For specific surface  $\Omega$-potential $\gamma$ we obtain 
\begin{equation}
\gamma(x') =  \int\limits_{-\infty}^{x'}\big [P(z') - P^{*}(x)\big ] dx +  \int\limits_{x'}^{+\infty}\big [P(z) - P^{*}(x)\big ] dx.
\label{eq:089}  
\end{equation}

Equation (\ref{eq:089}) generalizes the expression obtained previously \cite{StillingerBuff1962}, \cite{Zaskulnikov201102a} to the case of the finite value potentials. 

Certainly, expression (\ref{eq:042}) remains true in this case. 

Differentiating (\ref{eq:089}) with respect to the chemical potential and taking into account (\ref{eq:052}), we make sure that (\ref{eq:053}) is also true. 

From (\ref{eq:033}) taking into account (\ref{eq:009}) and (\ref{eq:086}) one can easily obtain an expression for the nonlinear surface coefficient 
\begin{equation}
\nu (z,z') =  \int\limits_{-\infty}^{\infty}\Big [ \widetilde{\theta}(x) P(z)+\widetilde{\varphi}(x)P(z')- P^*(x)  \Big ] dx
\label{eq:090},
\end{equation}
determining the form of $\nu$ in terms of ``local pressure''. The similarity of expressions (\ref{eq:090}) and (\ref{eq:079}) should be stressed.

\section{\label{subsec:03h}\texorpdfstring{``Mechanical definition'' of $\gamma$}{"Mechanical definition" of gamma}}%

The procedure of compression and extension of the system in two different directions with overall conservation of the volume and with changes of definite internal surfaces leads to this definition \cite[p. 43]{RowlinsonWidom2002}. This procedure will be applied in the upgraded form to the field/fluid interface. The consideration given below is a generalization of the procedure used for the case of infinite field \cite{Zaskulnikov201102a}.

In virtue of fundamental identity of the boundaries, it is necessary to take into account surface effects not only on the interface that we are interested in, but also at the boundaries of the system. 
\begin{figure}[htbp] 
  \includegraphics{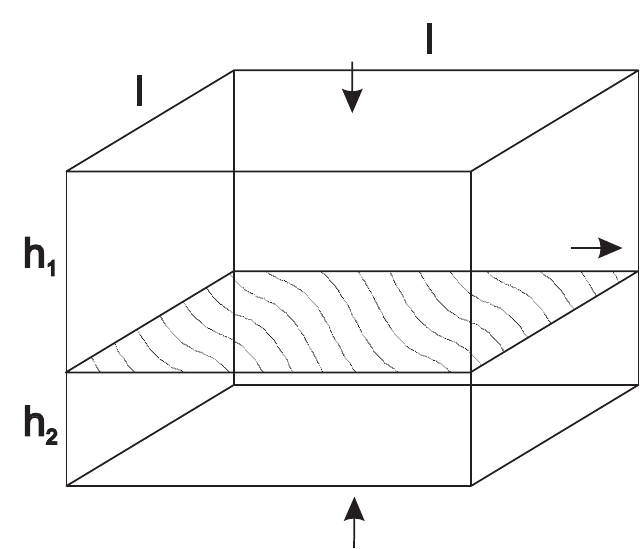}
   \caption{The system for the mechanical determination of the surface $\Omega$-potential of a fluid in a force field. The interface plane is singled out. The potential energies of the particles of the upper (index 1) and lower (index 2) parts are different.}
   \label{fig:01}
\end{figure}

We will consider the limiting physical surface to be inside the system at a microscopically large distance from its boundaries and at the same time this distance is much smaller than the size of the system. 

The procedure involves the following (see Fig.\ref{fig:01}). At first we move the upper plane of the system downward by $\delta h_1$ and the lower one by $\delta h_2$ upward. (These quantities are, obviously, the absolute values of displacements.) The work done by an external force on the system is
\begin{eqnarray}
\delta  W_1 =&& \Big [P_1(l - 2l_t)^2 + 4 l \int \limits_0^{l_t} P_{st_1} d x \Big] \delta h_1 \nonumber \\
&& + \Big [P_2(l - 2l_t)^2 + 4 l \int \limits_0^{l_t} P_{st_2} d x \Big] \delta h_2
\label{eq:091},
\end{eqnarray}
where $P_{st_i}$ is the pressure in the region of the narrow strips near the edges of the system \footnote[3]{For simplicity, we accept that all the lengths of the transition regions: $l_t, l_{st}, l^*, l_v$, connected respectively with $P_t, P_{st}, P^*, \theta$ are the same. One may consider on default for averaging integrals that the integration is performed over the maximal of these lengths.}, differing from pressure at the center (in the bulk). Index $s$ reminds that, generally speaking, this pressure does not coincide with the tangential component of pressure tensor \cite{Zaskulnikov201102a}
\begin{equation}
P_{st} \neq P_{t}
\label{eq:092}.
\end{equation} 

At the second stage, we allow the system to expand but in a lateral direction
\begin{eqnarray}
\delta W_2 =  -  \Bigl [ P_1&&(l - 2l_t)(h_1 - l_t) \nonumber \\
+ (l+2h_1)&& \int \limits_0^{l_t} P_{st_1} d x + P_2(l - 2 l_t)(h_2 - 2 l_t)\label{eq:093}   \\
&&  + (l+2h_2) \int \limits_0^{l_t} P_{st_2} d x+  l \int \limits_0^{l_t} P_{st} d x \Bigr ] \delta l.
\nonumber
\end{eqnarray}

Here $P_{st}$ is the tangential pressure along the boundary between the phases. The transition region is completely included into volume No. 2 (we speak here of the point separating heights $h_1$ and $h_2$).

Similarly to the case of infinite field, the condition for the conservation of volume is not quite trivial: we must consider the constancy of the volumes limited by the planes of arbitrary localization, set by coordinates $x$ for system boundaries and $x'$ for the phase interface ($x'$ will be directed upward from the origin of the transition region in the second phase). It is clear that each volume should be conserved constant separately in order to compensate the volume terms in the final expressions.

\begin{eqnarray}
&&  (l - 2x)^2 \delta h_1 = (l - 2x) (h_1 - x + l_t -x') \delta l \nonumber \\
&& (l - 2x)^2 \delta h_2 = (l - 2x) (h_2 - x - l_t + x') \delta l
\label{eq:094}
\end{eqnarray}
or conserving the terms of the first order of smallness,
\begin{eqnarray}
&& \delta h_1 = (h_1 + \frac{2h_1}{l}x-x+l_t-x')\frac{\delta l}{l}  \nonumber \\
&& \delta h_2 = (h_2 + \frac{2h_2}{l}x-x-l_t+x')\frac{\delta l}{l}
\label{eq:095}.
\end{eqnarray}

Then, in the same approximation, we obtain 
\begin{eqnarray}
\delta W = \delta W_1&& +\delta W_2 \nonumber\\
= \delta&& A_1 \Big [- \int \limits_0^{x} P_{st_1} dx + \int \limits_{x}^{l_t} (P_1 - P_{st_1}) dx \Big ] \nonumber \\
+ &&\delta A_2 \Big [- \int \limits_0^{x} P_{st_2} dx + \int \limits_{x}^{l_t} (P_2 - P_{st_2}) dx \Big ]
\label{eq:096} \\
&&+ \delta A \Big [ \int \limits_0^{x'}(P_2 - P_{st}) dx + \int \limits_{x'}^{l_t} (P_1 - P_{st}) dx \Big ], \nonumber 
\end{eqnarray}
where
\begin{equation}
\delta A_1 = (l - 2h_1) \delta l; ~~ \delta A_2 = (l - 2h_2) \delta l ~~ \text{и} ~~ \delta A = l \delta l 
\label{eq:097}
\end{equation}
are changes of the corresponding areas during the process. Similarly to the case of infinite field, the presence of free parameters ($h_1, h_2$) allows us further on to consider that only one surface is varied. 

In view of the assumptions made, the lower and upper integration limits may be considered to be equal to $-\infty$ and $+\infty$. So, it may be concluded that the expression for elementary work of change of the boundary surface area between the phases is 
\begin{equation}
\delta W =   \delta A \Big \{ \int \limits_{-\infty}^{x'} \big [P(z') - P_{st} \big ] dx + \int \limits_{x'}^{+\infty} \big [P(z) - P_{st}\big ] dx \Big \}.
\label{eq:098}
\end{equation}

Because the process takes place at constant $N$ and $T$, this work should be equal to the change of the free energy $F = F(V, T, N)$. An apparent complication during the process is pressure variation; generally speaking, pressure changes as a consequence of adsorption. Varying 
\begin{equation}
F = -P(z)(V-V') -P(z')V' + \gamma A + \mu (N_{b_1} + N_{b_2} + N_S) 
\label{eq:099}
\end{equation}
and taking into account the fact that $\delta N_{b_1} + \delta N_{b_2} = - \delta N_s$ and $\delta V = \delta V' = 0$ , we obtain
\begin{eqnarray}
\delta F =  -(V-&&V')\delta P(z) - V'\delta P(z')\label{eq:100}   \\
&& + \gamma \delta A + A \delta \gamma  +  (N_{b_1} + N_{b_2} + N_s) \delta \mu.
\nonumber
\end{eqnarray}

In the right side, all the terms are cancelled pairwise except the third term, so finally
\begin{equation}
\delta F =  \gamma \delta A 
\label{eq:101},
\end{equation}
as it must be if 
\begin{equation}
\gamma =    \int \limits_{-\infty}^{x'}\big [P(z') - P_{st}(x)\big ] dx + \int \limits_{x'}^{+\infty} \big [P(z) - P_{st}(x)\big ] dx.
\label{eq:102}
\end{equation}

So, the ``mechanical definition'' of $\gamma$ also comes to the structure that we already know (\ref{eq:081}), (\ref{eq:089}).

\section{\label{subsec:03i}\texorpdfstring{Tangential force and ``local pressure''}{Tangential force and "local pressure"}}%

Now we will demonstrate the consistency of approaches (\ref{eq:089}) and (\ref{eq:102}). We differentiate (\ref{eq:099}) with respect to the area, where we use for $\gamma$ (\ref{eq:089})
\begin{eqnarray}
\left ( \frac{\partial F}{\partial A} \right )_{N, V} && = -(V-V')\frac{\partial P(z)}{\partial A} -V'\frac{\partial P(z')}{\partial A} \nonumber \\
 + \gamma && + A \frac{\partial \gamma}{\partial A} + (N_s + N_{b_1} + N_{b_2}) \frac{\partial \mu}{\partial A}. 
\label{eq:103}
\end{eqnarray}

One can see that the first two terms in the right side of (\ref{eq:103}) are cancelled with $(N_{b_1} + N_{b_2})\partial \mu/\partial A$, and the fourth one with $N_s \partial \mu/\partial A$. We can easily make sure of this by passing under the integral to the derivative with respect to the chemical potential and taking into account (\ref{eq:052}), so that finally we obtain 
\begin{equation}
\left ( \frac{\partial F}{\partial A} \right )_{N, V} =  \int\limits_{-\infty}^{x'} [P(z') - P^{*}(x)] dx +  \int\limits_{x'}^{+\infty} [P(z) - P^{*}(x)] dx.
\label{eq:104}
\end{equation}

So, we may identify (\ref{eq:089}) and (\ref{eq:102}), and thus we obtain
\begin{equation}
\int\limits_{l_t}^{} P_{st} dx =   \int \limits_{l_t}^{} P^* dx
\label{eq:105}
\end{equation}
or
\begin{equation}
 \overline{P_{st}}  =     \overline{P^*},
\label{eq:106}
\end{equation}
where the bar means averaging over the transition layer. 

Under the real conditions, in the absence of clear boundaries of the transition layer, the accuracy of equality (\ref{eq:106}) is determined by the closeness of $P_{st}$ and $P^*$ to $P(z)$ and $P(z')$ at the boundaries of the averaging interval. The nontrivial part of the equality is determined exactly by the difference of these values from their limiting values. So, the value of averaging interval in (\ref{eq:106}) must determined by the compromise between these two factors.

\section{\label{subsec:03j}Pressure tensor and near-surface distortions}

Now we will prove a similar equality for the tangential component of the pressure tensor 
\begin{equation}
\overline{P_{t}} = \overline{P^*}.
\label{eq:107}
\end{equation}

We will rely on the equivalence of the problems of inhomogeneity in the field and inhomogeneity in the region of liquid/gas interface \cite{IrvingKirkwood1950}. The initial expression will be
\begin{equation}
P_{t}(x_1) = \varrho_1 k_B T  - \frac{1}{2} \int (y_{2} - y_{1}) \varrho^{(2)}_{12}  \frac{\partial u_{12}}{\partial y_2} d\bm{r}_2,
\label{eq:108}
\end{equation}
which is identical to that used in numerous works \cite{KirkwoodBuff1949, Harasima1958, onocondo1960, SchofieldHenderson1982}. Here $u_{ij}$ is the energy of the interaction of $i$-th and $j$-th particles, $y$ is the coordinate parallel to the surface, and the gradient of the external field is directed along $x$, similarly to the previous cases.

A brief outline of proof is as follows.

We integrate (\ref{eq:086}) and (\ref{eq:108}) over the coordinates of the first particle, then expand the subintegral expressions in the series of the activity using (\ref{eq:022}). Finally, equating the coefficients at the same powers of $z$, we prove their identity.

The equality of the integrals of $P_t$ and $P^*$ over volume is proved in Appendix \ref{sec:appendc}.

Now we will prove that $P_t = P^* = P$ in the region of homogeneity. Because all the directions are equivalent in this case, we may write down equations similar to (\ref{eq:108}) for $x$ and $z$. Summing these three equations we obtain 
\begin{eqnarray}
 3P_{t} &&= 3 \varrho k_B T  - \frac{1}{2} \int \varrho^{(2)}_{12} \Big [ (x_{2} - x_{1})  \frac{\partial u_{12}}{\partial x_2}  \nonumber\\
&&  + (y_{2} - y_{1})  \frac{\partial u_{12}}{\partial y_2} + (z_{2} - z_{1})  \frac{\partial u_{12}}{\partial z_2}  \Big ] d\bm{r}_2.
\label{eq:109}
\end{eqnarray}

Collapsing the bracketed expression we obtain a well known expression for pressure 
\begin{equation}
P_{t} =  \varrho k_B T  - \frac{1}{6} \int \varrho^{(2)}_{12} r_{12}  \frac{\partial u_{12}}{\partial r_{12}}  d\bm{r}_2 = P
\label{eq:110}
\end{equation}
\cite[p.190]{hillstatmeh1987}.

It is evident at once for $P^*$ that for $\theta_i = 1, ~i = 1,2, \dots$, (\ref{eq:086}) is reduced to (\ref{eq:009}), while for $\theta_i = \exp{(- \beta v_0)}$ to $P(z')$, that is, 
\begin{equation}
P^* = P(z)~ \text{or} ~P^* = P(z')
\label{eq:111}
\end{equation}
in the region of homogeneity.

So, in view of the validity of equalities (\ref{eq:110}), (\ref{eq:111}) in the region of homogeneity and the equality of the integrals of $P_t$ and $P^*$ over the system volume, we may conclude that (\ref{eq:107}) is true.

Similarly to the case of $P_{st}$, the region of the transition layer in (\ref{eq:107}) is determined by the closeness of $\theta$ to $\exp{(- \beta v_0)}$ and $1$ at the edges of the interval.

There are some reasons to suppose that the local equality is wrong, that is, generally speaking, 
\begin{equation}
 P_t  \neq    P^* ,
\label{eq:112}
\end{equation}
but a rigorous proof of (\ref{eq:112}) for the systems under consideration goes out of the scope of the present work. 

We will not consider also the questions connected with the ambiguity of pressure tensor because we are interested only in its zero moment \cite{SchofieldHenderson1982}.

So, in view of (\ref{eq:085}) and (\ref{eq:107}), we established a connection of the tangential force (per unit length) with the number density near the surface, and thus with the behavior of near-surface distortions 
\begin{equation}
\int \limits_{l_t}^{} \varrho(x) dx = \frac{\partial }{\partial \mu} \int \limits_{l_t}^{}P_t(x) dx
\label{eq:113}
\end{equation}
or
\begin{equation}
\overline{ \varrho} = \frac{\partial \overline{ P_t}}{\partial \mu},
\label{eq:114}
\end{equation}
where the correspondence to the macro-relation (\ref{eq:011}) also takes place.

At the same time, we obtained an equality for the tangential force acting on the transverse wall 
\begin{equation}
 \overline{P}_{st}  =   \overline{P}_t,
\label{eq:115}
\end{equation}
which cannot be considered self-evident in advance. 

Finally, expression (\ref{eq:089}) can be also written in the form 
\begin{equation}
\gamma(x') =  \int\limits_{-\infty}^{x'} \big [P(z') - P_{t}(x)\big ] dx +  \int\limits_{x'}^{+\infty} \big [P(z) - P_{t}(x) \big ] dx.
\label{eq:116}  
\end{equation}

This expression generalizes the Kirkwood-Buff equation for the arbitrary localization of the dividing surface \cite{Zaskulnikov201102a} to the case of the force field of finite value.

\section{\label{subsec:03k}Surface cluster expansion and pressure tensor approach}

Equality (\ref{eq:107}) allows us to prove the identity of cluster expansion and pressure tensor approaches in this general case too, similarly to the case of the impermeable wall \cite{Zaskulnikov201102a}. For this purpose, it is sufficient to demonstrate that (\ref{eq:043}) with $\nu$ in the form of (\ref{eq:033}) is equivalent to (\ref{eq:116}).

Equalizing these two expressions and canceling identical terms we obtain the condition of their identity in the form 
\begin{equation}
\nu (z,z') =  \int\limits_{-\infty}^{\infty}\Big [ \widetilde{\theta}(x) P(z)+\widetilde{\varphi}(x)P(z')- P_t(x)  \Big ] dx. 
\label{eq:117}
\end{equation}

Comparing (\ref{eq:117}) with (\ref{eq:090}) and taking into account (\ref{eq:107}) we ascertain that this equality is valid.

Taking into account (\ref{eq:071}), we obtained one more expression for the specific surface  $\Omega$-potential
\begin{eqnarray}
  \gamma && =  - \big[P(z)-P(z')\big](x'-x_0) \nonumber \\
&& + \int\limits_{-\infty}^{\infty}\Big [ \widetilde{\theta}(x) P(z)+\widetilde{\varphi}(x)P(z')- P_t(x)  \Big ] dx,
\label{eq:118}
\end{eqnarray}
of course, coinciding with expression (\ref{eq:116}) but again explicitly separating the terms linear and nonlinear with respect to pressure. Replacing $P_t$ in (\ref{eq:118}) by $P^*$ we return to the approach from the side of ``adsorption'', while replacing it by $P_{st}$  we come back to the ``mechanical definition'' of $\gamma$.

\section{\label{subsec:03l}Symmetry of the problem }

Unlike for the problem of adsorption when we have an impermeable wall, the problem of absorption is principally symmetrical. Varying the parameters we may consider one or another region of the fluid be free or situated in the field. 

Indeed, let us make the substitution   
\begin{eqnarray}
&& z^* = z'\nonumber \\
&& z'^* = z\nonumber \\
&& x^* = -x\label{eq:119} \\
&& v^*(x^*) = v(x) - v_0 \nonumber \\
&& v^*_0 = - v_0 \nonumber.
\end{eqnarray}

This set is equivalent to the original physical context, but mirrored with respect to $x =0$ and with a shift of the energy reference level.

Performing the substitution of (\ref{eq:119}) into (\ref{eq:026}), (\ref{eq:027}), we see that
\begin{eqnarray}
 \mkern -25mu\widetilde{\theta}[v(x),v_0] = \widetilde{\theta}[v^*(x^*)-v^*_0,-v^*_0]= \widetilde{\varphi}[v^*(x^*),v^*_0] && \nonumber \\
\mkern -25mu\widetilde{\varphi}[v(x),v_0] =  \widetilde{\varphi}[v^*(x^*)-v^*_0,-v^*_0] = \widetilde{\theta}[v^*(x^*),v^*_0] &&
\label{eq:120}.
\end{eqnarray}

So, as one can see, for example, from (\ref{eq:034}), (\ref{eq:035}) coefficient $\nu$ is invariant with respect to transformation (\ref{eq:119})
\begin{equation}
\nu(\widetilde{\theta},\widetilde{\varphi}|z,z') = \nu(\widetilde{\varphi}^*,\widetilde{\theta}^*|z'^*,z^*) =\nu(\widetilde{\theta}^*,\widetilde{\varphi}^*|z^*,z'^*) 
\label{eq:121},
\end{equation}
where we have introduced the notations
\begin{eqnarray}
\widetilde{\theta}^* =&& \widetilde{\theta}[v^*(x^*),v^*_0]  \nonumber \\
\widetilde{\varphi}^* =&&  \widetilde{\varphi}[v^*(x^*),v^*_0]
\label{eq:122},
\end{eqnarray}
and the dependence of $\nu$ on $\widetilde{\theta}, \widetilde{\varphi}$ is functional.

(In the derivation of (\ref{eq:121}), one must use the invariance of Ursell factors with respect to the change of coordinates sign.)

In turn, it follows from (\ref{eq:061}) that 
\begin{equation}
x^*_0 = - x_0
\label{eq:123},
\end{equation}
which, on substitution into (\ref{eq:071}) together with other parameters, gives 
\begin{equation}
\gamma(\widetilde{\theta},\widetilde{\varphi}|z, z',x') = \gamma(\widetilde{\theta}^*,\widetilde{\varphi}^*|z^*, z'^*,x'^*)
\label{eq:124}.
\end{equation}

Similarly, considering the surface number density (\ref{eq:072}) we obtain 
\begin{equation}
\varrho_s(\widetilde{\theta},\widetilde{\varphi},\varrho(x)|z, z',x') = \varrho_s(\widetilde{\theta}^*,\widetilde{\varphi}^*,\varrho^*(x^*)|z^*, z'^*,x'^*)
\label{eq:125},
\end{equation}
where $\varrho^*(x^*) = \varrho(x)$.

Finally, Henry constants are to be considered. It follows from the first two equations (\ref{eq:119}) that
\begin{eqnarray}
&& \varrho^* = \varrho'\nonumber \\
&& \varrho'^* = \varrho
\label{eq:126}.
\end{eqnarray}

Then, for Henry constant of absorption we have from (\ref{eq:062}) 
\begin{equation}
k_H^* = \frac{1}{k_H}
\label{eq:127},
\end{equation}
while for Henry constant of adsorption from (\ref{eq:056}) or (\ref{eq:057})
\begin{equation}
K_H^*(x'^*) z^* = K_H(x') z
\label{eq:128}.
\end{equation}

Another property determined by the symmetry has already been considered in section \ref{subsec:03b}.

Summing up, we may state that equations (\ref{eq:121}), (\ref{eq:124}) and (\ref{eq:125}) are control ones from the viewpoint of checking the problem on symmetry.

For small densities, the surface number density is determined by $\varrho_{s,l}$ (\ref{eq:073}). Demanding $\varrho_{s,l} > 0$, we see that for $\varrho(z) > \varrho(z')$ we must take $x'>x_0$.

So, if we accept that the denser medium is the fluid, it is the assignment of the transition layer to the solid body that agrees with the notion of adsorption as it is usually understood.

\section{\label{subsec:03m}High temperatures}

Using expression (\ref{eq:061}) and considering $\beta v(x) \ll 1$ within the entire range of $x$, we obtain
\begin{equation}
x_0 = - \int\limits_{-\infty}^{0} \Big [ 1 - \frac{v(x)}{v_0} \Big ] dx + \int\limits_{0}^{\infty} \frac{v(x)}{v_0} dx + \dots  
\label{eq:129}
\end{equation}

So, we see that $x_0$ is temperature-independent at high temperatures. It follows for Henry constant of adsorption from (\ref{eq:060}) that
\begin{equation}
K_H(x') = \beta v_0 (x' - x_0) + \dots = \beta \phi(x') + \dots
\label{eq:130},
\end{equation}
which corresponds to the limit (\ref{eq:065}) and where
\begin{equation}
\phi(x') =  \int\limits_{-\infty}^{x'} \big [ v_0 - v(x) \big ] dx - \int\limits_{x'}^{\infty} v(x) dx  
\label{eq:131}.
\end{equation}

The linear part of $\gamma$ is reduced to
\begin{eqnarray}
 - [P(z)&&-P(z')](x'-x_0)\label{eq:132}   \\
&& = - \varrho (z) v_0 (x'-x_0) + \dots = - \varrho (z) \phi(x') + \dots
\nonumber,
\end{eqnarray}
where we used the expansion of $P(z')$ in Taylor series near point $z$.

Then, for linear with respect to the density part of the surface number density (\ref{eq:075}) we have
\begin{equation}
\varrho_{s,l} = \varrho^2 v_0 (x' - x_0)\varkappa_T + \dots = \varrho^2 \varkappa_T \phi(x') + \dots
\label{eq:133},
\end{equation}
where 
\begin{equation}
\varkappa_T = \frac{1}{\varrho}\frac{\partial \varrho}{\partial P}
\label{eq:134}
\end{equation}
is isothermal compressibility. 

Now the nonlinear part of $\gamma$ will be calculated. For convenience of consideration, we will again turn to function 
\begin{equation}
{f}^{(k-1)}_{2...k}=\frac{{g}^{(k-1)}_{2...k}}{z^k}
\label{eq:135},
\end{equation}
and using (\ref{eq:035}) we obtain the form suitable for the given case 
\begin{eqnarray}
{f}^{(k-1)}_{2...k}= \int \limits_{- \infty}^{+ \infty}   \Big [ \theta(x) &- &\theta(x)\prod_{i = 2}^k \theta(x+x_i)   \label{eq:136}  \\
&  + & \varphi(x) \frac{\exp(-\beta v_0 k) - \exp (-\beta v_0)}{1 - \exp (-\beta v_0)} \Big ] d x
\nonumber.
\end{eqnarray}

Expanding the right side of (\ref{eq:136}) up to the terms of the second order with respect to $\beta u$ and making simple transformations we obtain 
\begin{equation}
{f}^{(k-1)}_{2...k} = \beta^2 \Big [ \sum_{i=2}^{k}\phi_{0i}    +  \sum_{i<j;i=2}^{j=k}\phi_{ij} \Big ] + \dots
\label{eq:137},
\end{equation}
where
\begin{equation}
\phi_{0i}  = \int \limits_{- \infty}^{+ \infty} v(x) [v_0 - v(x+x_i)] dx
\label{eq:138}
\end{equation}
and
\begin{equation}
\phi_{ij}  = \int \limits_{- \infty}^{+ \infty}  [v(x)v_0 - v(x+x_i)v(x+x_j)] dx
\label{eq:139}.
\end{equation}

Passing to $\nu$ with the help of (\ref{eq:034}), we have 
\begin{eqnarray}
\nu (z,T) &=&  \beta \sum_{k=2}^\infty \frac{z^k}{k!} \Big [ (k-1) \int \phi_{02} ~ {\cal U}^{(k)}_{0,2...k} d\bm{r}_2...d\bm{r}_k \label{eq:140} \\
&+& \frac{(k-1)(k-2)}{2} \int \phi_{23} ~ {\cal U}^{(k)}_{0,2...k} d\bm{r}_2...d\bm{r}_k \Big ] + \dots
\nonumber.
\end{eqnarray}

Then, differentiating (\ref{eq:140}) with respect to the chemical potential and summing the series, we obtain a connection between the nonlinear part of the surface number density and the first localized distribution functions
\begin{equation}
\varrho_{s,n} =  - \beta^2 \Big [ \int \phi_{02} ~ {\cal F}^{(2)}_{0,2} d\bm{r}_2 +  \frac{1}{2} \int \phi_{23} ~ {\cal F}^{(3)}_{0,2,3} d\bm{r}_2 d\bm{r}_3 \Big ] + \dots
\label{eq:141},
\end{equation}
where we used representation (\ref{eq:017}), as well as relation (\ref{eq:076}). It should be stressed that Ursell functions comprised in (\ref{eq:141}) relate to the uniform medium.

\section{\label{subsec:03n}Low temperatures}

Let $v_0> 0$, then it follows from (\ref{eq:025}) - (\ref{eq:027}) that for $T \to 0$ 
\begin{eqnarray}
&& z' \to 0 \nonumber \\
&& \widetilde{\theta}_i \to \theta_i   \label{eq:142} \\
&& \widetilde{\varphi}_i \to \varphi_i \nonumber.
\end{eqnarray}

In this situation, as one can see from (\ref{eq:061}), the coordinate of zero adsorption surface becomes 
\begin{equation}
x_0 = - \int\limits_{-\infty}^{0} \theta(x) dx + \int\limits_{0}^{\infty} \varphi(x) dx 
\label{eq:143},
\end{equation}
while Henry constant of adsorption is
\begin{equation}
K_H(x') = x' - x_0(T)
\label{eq:144},
\end{equation}
which follows from (\ref{eq:060}). For ${f}^{(k-1)}_{2...k}$, $\gamma$ and $\varrho_s$ we obtaine from (\ref{eq:136}), (\ref{eq:118}), (\ref{eq:072})  respectively,
\begin{equation}
{f}^{(k-1)}_{2...k}= \int \limits_{- \infty}^{+ \infty}  \theta(x) \Big [ 1 - \prod_{i = 2}^k \theta(x+x_i) \Big ] d x
\label{eq:145},
\end{equation}
\begin{equation}
\gamma =  - P(z)(x'-x_0) + \int\limits_{-\infty}^{\infty}\Big [ \theta(x) P(z)- P_t(x)  \Big ] dx
\label{eq:146},
\end{equation}
\begin{equation}
\varrho_s = \varrho(z)  (x'-x_0) + \int\limits_{-\infty}^{+\infty} \big [ \varrho(x) - \theta(x) \varrho(z) \big ]  dx
\label{eq:147}.
\end{equation}

Expressions (\ref{eq:143}) - (\ref{eq:147}), as expected, exactly correspond to the results for impermeable wall, obtained in \cite{Zaskulnikov201102a}.

\section{\label{subsec:03o}Static membranes}

Within the framework of the problem of permeable wall, the problem of static membrane can easily be solved. The static membrane will be understood as the region of time-independent field, which is a surface (that has, however, a certain thickness), with the field equal to zero on both sides of this region. In other words, the problem of permeable wall will be considered for
\begin{eqnarray}
&&  \beta v_0 = 0 \nonumber \\
&&  \beta v(x) \neq 0
\label{eq:148}.
\end{eqnarray}

The basic expressions can be most easily obtained from the forms like (\ref{eq:116}). Assuming $z' = z$ in (\ref{eq:052}), (\ref{eq:116}), we have 
\begin{equation}
\varrho_s =   \int\limits_{-\infty}^{\infty} \big [ \varrho(x) - \varrho \big ] dx 
\label{eq:149}
\end{equation}
and
\begin{equation}
\gamma =  \int\limits_{-\infty}^{\infty} \big [P - P_{t}(x)\big ] dx 
\label{eq:150},
\end{equation}
where $\varrho$ and $P$ are the number density and the pressure of the fluid far from the membrane. 

We obtain from (\ref{eq:042}) for this case
\begin{equation}
\Omega = -P V  + \gamma  A
\label{eq:151}.
\end{equation}

In the first glance, it may be concluded from (\ref{eq:151}) that there is no uncertainly in separating the volume and surface terms for membranes. In reality, for ``thick'' membranes, as one can easily understand from the form of $\gamma$ (\ref{eq:150}), this uncertainty is conserved: it is contained within $\gamma$. Indeed, the ``thick'' membrane does not differ from the voluminous field considered above. 

Henry constant of adsorption takes on the form
\begin{equation}
K_H =  \int\limits_{-\infty}^{\infty} \big [ \exp(-\beta v) - 1 \big ] dx 
\label{eq:152},
\end{equation}
which is most clearly seen from (\ref{eq:057}). 

In agreement with logics, $K_H$ is negative for positive potentials and vice versa.

If we start from the basic expression (\ref{eq:071}), we will have for the first term in the right side of this expression 
\begin{equation}
- \big [P(z)-P(z')\big ](x'-x_0) \rightarrow \varrho k_B T \int\limits_{-\infty}^{\infty} \big [ 1 - \exp(-\beta v) \big ] dx 
\label{eq:153}.
\end{equation}

The term with $x'$ becomes zero, which is logical: the disappearance of phases must mean the disappearance of interface. The value of $x_0 \rightarrow \infty$: in this case adsorption is always present. 

The nonlinear surface coefficient becomes in this case 
\begin{equation}
\nu(z,z) =  \int\limits_{-\infty}^{\infty} \big [P - P_{t}(x)\big ] dx - \varrho k_B T \int\limits_{-\infty}^{\infty} \big [ 1 - \exp(-\beta v) \big ] dx 
\label{eq:154},
\end{equation}
which can be obtained, for example, through expansion of (\ref{eq:117}) in series of $\beta v_0$.

Because equations (\ref{eq:106}), (\ref{eq:107}) remain true for this case, the expressions similar to the above-mentioned ones with the substitutions $P_t \rightarrow P^*$ and $P_t \rightarrow P_{st}$ are valid too.

At last, expression for the function ${f}^{(k-1)}$ takes the form
\begin{equation}
{f}^{(k-1)}_{2...k}= \int \limits_{- \infty}^{+ \infty}   \Big [1 - k\varphi(x)  - \theta(x)\prod_{i = 2}^k \theta(x+x_i) \Big ] d x
\label{eq:155},
\end{equation}
which follows from the expansion of (\ref{eq:136}) in series of $\beta v_0$. Replacement of (\ref{eq:155}) into (\ref{eq:034}) gives (\ref{eq:154}) for the version with $P^*$, if the order of integration is changed.

\section{\label{subsec:03p}Near-surface virial expansion}

Equation (\ref{eq:114}) and the properties of (\ref{eq:107}), (\ref{eq:115}) allow us to construct the near-surface analogs of virial expansion. The results of this building coincide with the results of \cite{Zaskulnikov201102a} for an impermeable wall.

For the case of the uniform medium we have the system of equations (\ref{eq:009}), (\ref{eq:010}), and it may be written in the form  
\begin{equation}
	 \left\{ 
			\begin{array}{cll} 			
        \beta P &=&\displaystyle{z + \sum_{n=2}^{\infty} b_n z^n}  \\   
         \varrho &=& \displaystyle{z + \sum_{n=2}^{\infty} n b_n z^n} 
     	\end{array}  
		\right.
		\label{eq:156},
\end{equation}
where
\begin{equation}
b_n = \frac{1}{n!} \int {\cal U}^{(n)}_{1...n} d\bm{r}_2...d\bm{r}_n
\label{eq:157}.
\end{equation}

It should be stressed that integration in (\ref{eq:157}) is performed in infinite limits \cite{zaskulnikov200911a}.

Excluding $z$ we obtain 
\begin{equation}
\beta P = \varrho - \sum_{k=1}^{\infty} \frac{k}{k+1} \beta_k \varrho^{k+1}
\label{eq:158},
\end{equation}
where
\begin{equation}
\beta_k = \sum_{\bm{m}} (-1)^{\sum_j m_j -1} \frac{(k-1+\sum_j m_j)!}{k!} \prod_j \frac{(jb_j)^{m_j}}{m_j!}
\label{eq:159}
\end{equation}
are so-called irreducible cluster integrals. Summing is performed over all sets of nonnegative $m_j$ meeting the requirement 
\begin{equation}
\sum_{j=2}^{k+1} (j-1) m_j = k
\label{eq:160}
\end{equation}
\cite[p. 144]{hillstatmeh1987}, which provides the same dimensionality of the terms of sum (\ref{eq:159}).

In the presence of the field, we must start from equations (\ref{eq:086}), (\ref{eq:020}). Averaging them over the transition layer we obtain the system 
\begin{equation}
	 \left\{ 
			\begin{array}{cll} 			
        \beta \overline{P^*}  &=&\displaystyle{ z \overline{\theta} + \sum_{n=2}^{\infty} d_n (z\overline{\theta})^n}  \\   
         \overline{ \varrho} &=& \displaystyle{z \overline{\theta} + \sum_{n=2}^{\infty} n d_n (z\overline{\theta})^n} 
     	\end{array}  
		\right.
		\label{eq:161},
\end{equation}
where
\begin{equation}
d_n = \frac{1}{n!\, l_t \,\overline{\theta}^n} \int \limits_{l_t}  \int \Big [ \prod_{i = 1}^{n} \theta_i \Big ] {\cal U}^{(n)}_{1...n} d x_1 d\bm{r}_2...d\bm{r}_n
\label{eq:162}.
\end{equation}

Here integration over the coordinate of $1$-th particle is made in the limits of the transition layer, while integration over the coordinates of other particles is performed in infinite limits.

Similarly to (\ref{eq:033}) we may integrate over $x_1$. Making the change of variables $x'_1 = x_1,~\bm{r}'_i = \bm{r}_i -\bm {r}_1$ ($i = 2,\dots n$), using the invariance of ${\cal U}^{(k)}_{1...k}$ with respect to translations and dropping the primes, we get 
\begin{equation}
d_n = \frac{1}{n!} \int  {h}^{(n-1)}_{2...n} {\cal U}^{(n)}_{0,2...n} d\bm{r}_2...d\bm{r}_n
\label{eq:163},
\end{equation}
where
\begin{equation}
{h}^{(n-1)}_{2...n} = \frac{1}{l_t\,\overline{\theta}^n} \int \limits_{l_t} \theta(x)  \Big [ \prod_{i = 2}^n\theta(x + x_i) \Big ] d x
\label{eq:164}.
\end{equation}

In view of the identity of the functional connection of $P$ and $\varrho$ in (\ref{eq:156}) on the one hand, and $\overline{P^*}$ and $\overline{ \varrho}$ in (\ref{eq:161}) on the other hand, we can immediately write the analog of (\ref{eq:158})
\begin{equation}
\beta \overline{P^*} = \overline{ \varrho} - \sum_{k=1}^{\infty} \frac{k}{k+1} \delta_k \overline{ \varrho}^{k+1}
\label{eq:165},
\end{equation}
where
\begin{equation}
\delta_k = \sum_{\bm{m}} (-1)^{\sum_j m_j -1} \frac{(k-1+\sum_j m_j)!}{k!} \prod_j \frac{(jd_j)^{m_j}}{m_j!} 
\label{eq:166},
\end{equation}
while requirement (\ref{eq:160}) is conserved in the previous form.

Finally, taking into account (\ref{eq:107}), (\ref{eq:115}), we may write
\begin{equation}
\beta \overline{P}_t = \overline{\varrho} - \sum_{k=1}^{\infty} \frac{k}{k+1} \delta_k \overline{ \varrho}^{k+1}
\label{eq:167}
\end{equation}
and
\begin{equation}
\beta \overline{P}_{st} = \overline{\varrho} - \sum_{k=1}^{\infty} \frac{k}{k+1} \delta_k \overline{ \varrho}^{k+1}
\label{eq:168}.
\end{equation}

Expressions (\ref{eq:167}) and (\ref{eq:168}) are preferable than (\ref{eq:165}): both $P_t$ and $P_{st}$ can be derived from computer experiments. In addition, $P_t$ can be calculated because it may be represented in quadratures (\ref{eq:108}).

From the viewpoint of the nontriviality of equalities, averaging intervals in (\ref{eq:167}), (\ref{eq:168}) are to be chosen minimal but such that the equalities (\ref{eq:106}), (\ref{eq:107}) would still be valid.

Thus, a connection of $\delta_k$ with $d_j$ is determined by equation (\ref{eq:166}). For example, several first relations are
\begin{eqnarray}
&& \delta_1 = 2 d_2  \nonumber \\
&& \delta_2 = 3 d_3 - 6 d_2^2 \label{eq:169} \\
&& \delta_3 = 4 d_4 -24 d_2 d_3 + \frac{80}{3} d_2^3  \nonumber \\
&& \dots 
\nonumber,
\end{eqnarray}
and they are identical to the corresponding relations between $\beta_k$ and $b_j$ \cite[p. 144]{hillstatmeh1987}.

\section{\label{sec:04}Summary}

\begin{enumerate}

	\item The general expression was obtained for the $\Omega$-potential (grand potential) of the system into which a force field of a finite value was locally introduced (\ref{eq:032}). This approach allows one to introduce an arbitrary position of the separating surface (\ref{eq:036}). 	
	\item The volume and surface terms of the $\Omega$-potential cannot be separated unambiguously; their interpretation is determined by the shape of the interaction potential between the body and the particles of the fluid (\ref{eq:037}) - (\ref{eq:039}). 		
	\item In particular, depending on the situation, the surface terms can start from a term which is either linear or quadratic with respect to the activity (\ref{eq:037}), (\ref{eq:039}), which corresponds to the presence or absence of adsorption in the classical understanding of this term. 	
	\item The general expression was obtained for specific surface  $\Omega$-potential of the nonuniform system ($\gamma$), it is composed of the parts that are linear and nonlinear with respect to pressure (\ref{eq:071}). The linear part is determined by the product of pressure difference and the distance from the separating surface to the surface of zero adsorption (\ref{eq:059}), while the nonlinear part (the nonlinear surface  coefficient - $\nu$) has several universal representations (\ref{eq:033}), (\ref{eq:034}), (\ref{eq:090}), (\ref{eq:117}).	
	\item The surface number density, similarly to $\gamma$, is separated into the part linear with respect to density, which depends on the position of dividing surface, and a universal nonlinear one (\ref{eq:072}).	
	\item Henry constant of adsorption depends on the position of the dividing surface and is determined by expression (\ref{eq:060}). 	
	\item An expression for the nonlinear surface coefficient $\nu$ was obtained in the form of the series in powers of the activity, which is an analog of the cluster expansion for pressure (\ref{eq:034}). The coefficients of the series are represented as the integrals of the products of Ursell factors and multipliers depending on the external potential (\ref{eq:035}). 	
	\item For macroscopically smooth field of arbitrary configuration (section \ref{subsec:03f}) the expression obtained for $\nu$ (\ref{eq:079}) agrees with the expression for the  $\Omega$-potential of the nonuniform system (\ref{eq:080}).		
	\item The approach through pressure tensor gives an expression for $\nu$ in quadratures (\ref{eq:117}), which allows one to operate on $\gamma$ in whole (\ref{eq:118}).	
	\item The approach through the cluster expansion gives the expression for $\nu$ in quadratures when using the  ``local pressure'' (\ref{eq:090}).	
	\item As an average over the transition region in the vicinity of the surface, the tangential component of the pressure tensor (\ref{eq:108}) coincides with the pressure acting on the transverse wall (\ref{eq:091}) and with the ``local pressure'' (\ref{eq:086}). This fact, as well as the arbitrary position of the dividing surface, allows us to consider the new variants of Kirkwood-Buff formula (\ref{eq:089}), (\ref{eq:102}), (\ref{eq:116}). 		
	\item The  $\Omega$-potential of the system in the field of arbitary configuration is determined by the integral of ``local pressure'' over the volume (\ref{eq:083}), (\ref{eq:086}). In view of the previous item, it is also determined by analogous integrals of $P_t$ and $P_{st}$, when these values may be introduced.		
	\item As an average over the transition region in the vicinity of the surface, the relation between the number density and the tangential component of pressure tensor corresponds to the usual macroscopic relation between a number density and a pressure (\ref{eq:114}).	
	\item Approaches through the cluster expansion and through the pressure tensor for the problems under consideration are completely equivalent within the domain of their existence (section \ref{subsec:03k}).		
	\item The solution of the problem of interface in the force field of a finite value possesses the symmetry with respect to the permutation of the field and the free fluid (section \ref{subsec:03l}).	
	\item Within the problem under consideration, expressions for $\gamma $, $K_H$, $\varrho_S$ were obtained for static membranes (section \ref{subsec:03o}).	
	\item Within the problem under consideration, the parametric relation between Henry constants of adsorption and absorption was obtained (section \ref{subsec:03d}).	
	\item At high temperatures, the nonlinear part of the surface number density is expressed through the functionals of Ursell functions of the 2-nd and 3-rd order (\ref{eq:141}).	
	\item The near-surface virial expansion (section \ref{subsec:03p}) gives the equation of state for the ``two-dimensional'' fluid in the region near the boundary.

\end{enumerate}

\begin{appendices}

\numberwithin{equation}{section}

\section{\label{sec:appenda}Factors}

\subsection{\label{subsec:appenda1}Ursell factors }

Ursell factors can be defined by equality:
\begin{eqnarray}
{\cal U}^{(k)}_{1...k} &=& \sum_{\{\bm{n}\}}(-1)^{l-1}(l-1)!\prod_{\alpha = 1}^l \exp(-\beta U^{k_\alpha}(\{\bm{n}_\alpha\})),  \nonumber \\
1 & \leq & ~ k_\alpha \leq k, ~~ \sum_{\alpha = 1}^l k_\alpha = k, ~ \exp(-\beta U^{1}) = 1,
\label{eq:a01}
\end{eqnarray}
where $\{\bm{n}\}$ designates some partition of the given set of $k$ particles with coordinates $\bm{r}_1,\dots,\bm{r}_k$  into disjoint groups $\{\bm{n}_\alpha\}$, $l$ is the number of groups of given partition, $k_\alpha$ is the size of the group with the number $\alpha$. Summing is made over all the possible partitions; the sense of the requirement $\exp(-\beta U^{1}) = 1$ is evident: groups containing a sole particle do not give any contribution into the product in this case. 

For example, some initial ${\cal U}^{(k)}_{1...k}$  are
\begin{eqnarray}
{\cal U}^{(1)}_{1}\mkern 9mu &=& 1 \nonumber\\
{\cal U}^{(2)}_{1,2} \mkern 8mu &=& \exp(-\beta U^{2}_{1,2}) - 1 \label{eq:a02}  \\
{\cal U}^{(3)}_{1,2,3} &=& \exp(-\beta U^{3}_{1,2,3}) - \exp(-\beta U^{2}_{1,2}) \nonumber \\
 &-& \exp(-\beta U^{2}_{1,3}) - \exp(-\beta U^{2}_{2,3}) + 2 \nonumber \\
\dots \nonumber
\end{eqnarray}

It is well known that one of the generating functions for Ursell factors is logarithm \cite{percus1964}. However, dividing the series in (\ref{eq:008}) for the case of $k=1$ by $\Xi_V$ in the form of (\ref{eq:006}), one can easily make sure that there is another generating function - a fractional function, in the form of the ratio of series (\ref{eq:008}). This follows from the well known expansion for the number density of the uniform system (\ref{eq:010}), which is true far from the boundaries of the system.

The existence of the fractional generating function will be proved in Appendix \ref{subsec:appendb1}.

\subsection{\label{subsec:appenda2}Factors of partial localization}

As far as we know, these functions were introduced for the first time in \cite{RuelleStatmeh1969}. 

Some particles involved in these functions do not cause the decay when moving away (the delocalized group), and some of them - do (the localized one). 

Introduce the notation
\begin{equation}
{\cal B}^{(m,k)}_{1...m+k},
\label{eq:a03}
\end{equation}
where the superscripts $m$ and $k$ define the quantity of delocalized and localized particles, respectively ($m = 1,2,3,\dots, k = 0,1,2,\dots$). The subscripts denote the coordinates of particles, with the first $m$ particles being considered delocalized, and the rest - localized.

These functions are similar in structure to the Ursell factors of  $k+1$-th rank, with the proviso that in the construction by the type of (\ref{eq:a01}) the first $m$ particles (delocalized) are treated as a single compound particle. In other words, define ${\cal B}^{(m,k)}_{1...m+k}$ by the equality  
\begin{eqnarray}
{\cal B}^{(m,k)}_{1...m+k} &=& \sum_{\{\bm{n}\}}(-1)^{l-1}(l-1)!\nonumber \\
&\times& \prod_{\alpha = 1}^l \exp(-\beta U^{k_\alpha+(m-1)\delta_{\alpha\tau}}(\{\bm{n}_\alpha\})), \label{eq:a04} \\
1  \leq&  k_\alpha& \leq k+1; ~~ \sum_{\alpha = 1}^l k_\alpha = k+1; ~~ \exp(-\beta U^{1}) = 1,\nonumber 
\end{eqnarray}
where designations are analogous to (\ref{eq:a01}) on condition that summation is taken over all possible partitions of the set of $k+1$ particles among which one particle is compound. $\delta_{\alpha\tau}$ is the Kronecker delta, and $\tau$ is the number of the group involving a compound particle. 

The generating function for ${\cal B}^{(m,k)}_{1...m+k}$  is the distribution function $\varrho^{(m)}_{G,1...m}$ of GCE type. Expanding the partition function $\Xi_V$ in (\ref{eq:008}) and dividing the series, we obtain (\ref{eq:013}). The proof of this relation is given in Appendix \ref{subsec:appendb1}. 

The first in localized group ${\cal B}^{(m,k)}_{1...m+k}$ are 
\begin{eqnarray}
{\cal B}^{(m,0)}_{1...m} \mkern 12mu &=& \exp(-\beta U^{m}_{1...m}) \nonumber\\
{\cal B}^{(m,1)}_{1...m+1} &=& \exp(-\beta U^{m+1}_{1...m+1}) - \exp(-\beta U^{m}_{1...m})\label{eq:a05} \\
{\cal B}^{(m,2)}_{1...m+2} & = & \exp(-\beta U^{m+2}_{1...m+2}) -  \exp(-\beta U^{m+1}_{1...m+1})\nonumber \\
 &-& \exp(-\beta U^{m+1}_{1...m,m+2}) - \exp(-\beta U^{m}_{1...m}) \nonumber \\
 &\times& \exp(-\beta U^{2}_{m+1,m+2})  + 2\exp(-\beta U^{m}_{1...m}) \nonumber \\
\dotso \nonumber
\end{eqnarray}
and in delocalized one are
\begin{equation}
{\cal B}^{(1,k-1)}_{1...k} = {\cal U}^{(k)}_{1...k}
\label{eq:a06},
\end{equation}
including, for homogeneous medium 
\begin{equation}
{\cal B}^{(1,0)}_{1} = 1
\label{eq:a07}.
\end{equation}

It is obvious from (\ref{eq:a05}), and (\ref{eq:a06}) that factors of partial localization generalize the concepts of Boltzmann factors and Ursell factors, including them as the limiting cases.  

For factors ${\cal B}^{(m,k)}_{1...m+k}$ a number of recurrence relations hold, which ensure the existence of various physical links; some of them are given in Appendix \ref{sec:appendb}.

\section{\label{sec:appendb}\texorpdfstring{Recurrence relations for ${\cal B}^{(m,k)}_{1...m+k}$}{Recurrence relations for B(m,k)1...m+k}}

Many relations and operations of statistical mechanics are provided by definite classes of recurrence relations for ${\cal B}^{(m,k)}_{1...m+k}$. For brevity, we shall say that an operation generates a recurrence relation or a class. In this contribution only some of them will be considered. More detailed information can be found in \cite{zaskulnikov201004a}.

\subsection{\label{subsec:appendb1}Correspondence to the definition}

Equating expressions (\ref{eq:008}) and (\ref{eq:013}), expanding the series for $\Xi_V$, and performing multiplication of the series, we arrive at 
\begin{equation}
{\cal B}^{(m,k)}_{1...m+k} = {\cal B}^{(m+k,0)}_{1...m+k} - \sum_{n=1}^k \sum_{\text{samp}} {\cal B}^{(n,0)}_{1...n}{\cal B}^{(m,k-n)}_{n+1...m+k},
\label{eq:b01}
\end{equation}
where $m \geq 1$, $k \geq 1$, and the internal sum is taken over samplings $n$ of initial localized $k$ particles. Relation (\ref{eq:b01}) is proved either by direct exhaustion of partitions in accordance with (\ref{eq:a04}), or by repeatedly substitutions the expression for ${\cal B}^{(m,k)}_{1...m+k}$ in the right-hand side of (\ref{eq:b01}). 

Paper \cite{percus1964}  gives another recurrence relation for Ursell functions which in terms of ${\cal B}^{(m,k)}_{1...m+k}$ looks like
\begin{equation}
{\cal B}^{(1,k)}_{1...k+1} = {\cal B}^{(k+1,0)}_{1...k+1} - \sum_{n=1}^{k} \binom{k}{n}\left [ {\cal B}^{(n,0)}_{1...n} {\cal B}^{(1,k-n)}_{n+1...k+1}\right ]_{\text{perm}},
\label{eq:b02}
\end{equation}
where $k \geq 1$, and square brackets denote averaging over permutations of particles. 

In principle, this is the same equation (\ref{eq:b01}) at $m=1$ but written in a symmetric form. Note that in the given case there is no necessity in symmetrization - equation  (\ref{eq:b01}) holds rigorously in asymmetric form as well. Nevertheless, (\ref{eq:b01}) can also be brought into a symmetric form, however, this calls for averaging over permutations of both the sums and the left-hand side of the equation.

\subsection{\label{subsec:appendb2}BBGKI equation}

In the presence of external field this set of linking equations may be written as \cite[p.205]{hillstatmeh1987}
\begin{eqnarray}
{\nabla}_1\varrho^{(m)}_{1...m} = -&& \beta  \varrho^{(m)}_{1...m} \Big({\nabla}_1 v_1 + \sum_{i=2}^{m }{\nabla}_1 u_{1i} \Big ) \nonumber \\ && - \beta  \int  \varrho^{(m+1)}_{1...m+1} {\nabla}_1 u_{1m+1} d\bm{r}_{m+1}. \label{eq:b03}
\end{eqnarray}

This relation, after substituting expressions (\ref{eq:022}) in it and equating coefficients at equal powers of $z$, generates the differential recurrence relation
\begin{eqnarray}
{\nabla}_1{\cal B}^{(m,n)}_{1...m+n} = -&& \beta  {\cal B}^{(m,n)}_{1...m+n} \sum_{i=2}^{m }{\nabla}_1 u_{1i} \nonumber \\ -&& \beta \sum_{i=m+1}^{m+n }{\cal B}^{(m+1,n-1)}_{1i..m+n}{\nabla}_1 u_{1i},
\label{eq:b04}
\end{eqnarray}
which we use to prove (\ref{eq:107}), (\ref{eq:113}). In the second term on the right all permutations of particles of the localized group ($n$) with the second particle are exhausted.

\section{\label{sec:appendc}\texorpdfstring{Proof of $\langle P_t\rangle = \langle P^*\rangle$}{Proof of <Pt> = <P*>}}

\begin{figure}[htbp]   
  \includegraphics{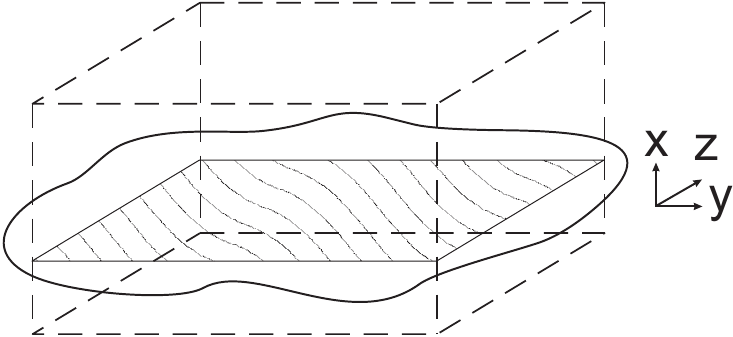}
   \caption{Система, используемая для доказательства $\langle P_t\rangle = \langle P^*\rangle$.}
   \label{fig:02}
\end{figure}

Let us prove the equality of the integrals of $P^*$ and $P_t$ taken over the entire volume. To prove this equality, consider the system inside the volume shaped as rectangular parallelepiped strongly stretched along the $y$ axis (see Fig.\ref{fig:02}). The interface is inside the system. The significance of field/fluid interaction is still defined by the functions $\theta$.

Let us integrate expressions (\ref{eq:086}) and (\ref{eq:108}) with respect to the coordinates of the first particle. The integration is performed over the whole system volume denoted by dashed line in Fig. \ref{fig:02}. We put that this volume is defined by the characteristic function $\psi_1$.  

Expand the integrands into a series of the activity using (\ref{eq:022}). Equating the coefficients at equal powers of $z$, we have 
\begin{eqnarray}
&&  k_B T \int \psi_1 \Big [ \prod_{i = 1}^{k+2} \theta_i \Big ] {\cal B}^{(1,k+1)}_{1...k+2} d\bm{r}_1 ...d\bm{r}_{k+2} \label{eq:c01}\\
&& = \frac{(k+2)}{2} \int \psi_1 \Big [ \prod_{i = 1}^{k+2} \theta_i \Big ] (y_2 - y_1) {\cal B}^{(2,k)}_{1...k+2} \frac{\partial u_{12}}{\partial y_2} d\bm{r}_1 ...d\bm{r}_{k+2},
\nonumber 
\end{eqnarray}
where $k = 0,1,\dots$. Here we equated the coefficients at $z^{k+2}$. The kind of the recurrence relation (\ref{eq:b04}) required to prove (\ref{eq:c01}) is of the form 
\begin{equation}
- k_B T \frac{\partial}{\partial y_2}{\cal B}^{(1,k+1)}_{1...k+2} = \sum_{i=1,i\neq2}^{k+2 }{\cal B}^{(2,k)}_{2i..k+2}\frac{\partial u_{2i}}{\partial y_2} .
\label{eq:c02}
\end{equation}

Multiplying (\ref{eq:c02}) by $(y_2 - y_1)\psi_1\theta_1\dots\theta_{k+2}$ and integrating over all coordinates, we obtain using the symmetry under permutations inside localized and delocalized groups of ${\cal B}^{(m,n)}$ 
\begin{eqnarray}
 - k_B T  \int \psi_1 \Big [ \prod_{i = 1}^{k+2} \theta_i \Big ] (y_2 - y_1)&& \frac{\partial}{\partial y_2}{\cal B}^{(1,k+1)}_{1...k+2} d\bm{r}_1 ...d\bm{r}_{k+2} \nonumber \\
&& =  I_1 + k I_2 +k I_3, \label{eq:c03} 
\end{eqnarray}
where
\begin{equation}
I_1=\int  \psi_1  \Big [ \prod_{i = 1}^{k+2} \theta_i \Big ] (y_2 - y_1) {\cal B}^{(2,k)}_{2,1...k+2} \frac{\partial u_{21}}{\partial y_2} d\bm{r}_1 ...d\bm{r}_{k+2},
\label{eq:c04}
\end{equation}
\begin{equation}
I_2 = \int \psi_1 \Big [ \prod_{i = 1}^{k+2} \theta_i \Big ] (y_2 - y_3) {\cal B}^{(2,k)}_{2,1...k+2} \frac{\partial u_{21}}{\partial y_2} d\bm{r}_1 ...d\bm{r}_{k+2},
\label{eq:c05}
\end{equation}
\begin{equation}
I_3 = \int (\psi_3 - \psi_1)\Big [ \prod_{i = 1}^{k+2} \theta_i \Big ] (y_2 - y_3) {\cal B}^{(2,k)}_{2,1...k+2} \frac{\partial u_{21}}{\partial y_2} d\bm{r}_1 ...d\bm{r}_{k+2}.
\label{eq:c06}
\end{equation}

Here in the integrals $I_2$ and $I_3$ the change of variables $\bm{r}_1 \leftrightarrow \bm{r}_3$ is performed. 

The integral in the left-hand side of equality (\ref{eq:c03}) may be taken by parts, since here factors $\theta_i$ depend solely on $x_i$. Wherein, due to locality of Ursell factors, and factors ${\cal B}^{(1,n)}$ are exactly the factors of this kind, the integral term goes to zero. Thus, 
\begin{eqnarray}
 - k_B T && \int  \psi_1 \Big [ \prod_{i = 1}^{k+2} \theta_i \Big ] (y_2 - y_1) \frac{\partial}{\partial y_2}{\cal B}^{(1,k+1)}_{1...k+2} d\bm{r}_1 ...d\bm{r}_{k+2} \nonumber \\
&& = k_B T \int \psi_1 \Big [ \prod_{i = 1}^{k+2} \theta_i \Big ] {\cal B}^{(1,k+1)}_{1...k+2} d\bm{r}_1 ...d\bm{r}_{k+2} \label{eq:c07}. 
\end{eqnarray}

It is readily seen that 
\begin{equation}
I_1 = 2 I_2.
\label{eq:c08}
\end{equation}

For this purpose, calculate the difference
\begin{equation}
I_1 - I_2 = \int  \psi_1  \Big [ \prod_{i = 1}^{k+2} \theta_i \Big ] (y_3 - y_1) {\cal B}^{(2,k)}_{2,1...k+2} \frac{\partial u_{21}}{\partial y_2} d\bm{r}_1 ...d\bm{r}_{k+2}.
\label{eq:c09}
\end{equation}

Making the change of variables $\bm{r}_1 \leftrightarrow \bm{r}_2$, we see that the right-hand side of (\ref{eq:c09}) coincides with $I_2$, so we have (\ref{eq:c08}). 

Finally, consider the integral $I_3$. Obviously, the factor $(\psi_3 - \psi_1)$ is nonzero only in the case when the first and the third particles are on different sides of the system boundary. It is seen better from the expression
\begin{equation}
\psi_3 - \psi_1 = \psi_3\chi_1 - \psi_1\chi_3,
\label{eq:c10}
\end{equation}
where we introduce the function $\chi_i =  1 - \psi_i$.

When the $1$-st or the $3$-rd particle moves away from the boundary of the system, the integrand decays rapidly owing to local character of the product ${\cal B}^{(2,k)}_{2,1...k+2}\partial u_{21}/\partial y_2$. Hence, the integral is proportional to the surface of the parallelepiped at hand, but not all of its faces make a contribution to $I_3$.

Consider the top face of the system, where due to remoteness from the interface we can consider that $\theta_i = 1, i = 1,2,3\dots$. Making the substitution (\ref{eq:c10}), we arrive at the conclusion that condition $I_3 = 0$ demands in this case that
\begin{eqnarray}
\int \psi_3\chi_1 &&(y_2 - y_3) {\cal B}^{(2,k)}_{2,1...k+2} \frac{\partial u_{21}}{\partial y_2} d\bm{r}_1 ...d\bm{r}_{k+2} \label{eq:c11} \\
&&= \int \psi_1\chi_3 (y_2 - y_3) {\cal B}^{(2,k)}_{2,1...k+2} \frac{\partial u_{21}}{\partial y_2} d\bm{r}_1 ...d\bm{r}_{k+2}.
\nonumber
\end{eqnarray}

However, obviously, equation (\ref{eq:c11}) is true by symmetry when the $y$ axis is parallel to the plane defined by the function $\psi$. To prove this it suffices to consider the mirror reflection operation with its mirror-plane coinciding with the upper face of the system. Thus, we may conclude that contribution to $I_3$ from the whole region of homogeneity above the interface equal zero except the side faces $y = const$. In the region of homogeneity below the interface situation is completely analogous due to the fact that in this region $\theta_i = const, i = 1,2,3\dots$.

So, only the side faces $y = const$ and two narrow strips on the faces $z = const$ make contributions to $I_3$. The contribution of these strips is of higher order of smallness (proportional the linear sizes of the system) and may be neglected. Thus, we have the estimation
\begin{equation}
\frac{I_3}{S_{zy}} =  o(S_{zy}).
\label{eq:c12}
\end{equation}

The contribution of $I_3$ tends to zero with the system length (along the $y$ axis) tending to infinity. Here $S_{zy}$ is the area of the face $zy$. 

In view of (\ref{eq:c08}) and (\ref{eq:c12}), equation (\ref{eq:c03}) gives (\ref{eq:c01}).

\end{appendices}

\begin{strip}
\center{---------------------------------------------------------------------------------------------}
\end{strip}

\tiny 
\raggedleft
VZ, 08.12.2011, v041.


\begin{thebibliography}{10}

\bibitem{Ono1950}
S.~Ono.
\newblock Application of ursell and mayer's treatment for imperfect gases to
  adsorption.
\newblock {\em J. Chem. Phys.}, 18:397, 1950.

\bibitem{hill1959}
T.~L. Hill.
\newblock Relations between different definitions of physical adsorption.
\newblock {\em J. Phys. Chem.}, 63(4):456--460, 1959.

\bibitem{onocondo1960}
S.~Ono and S.~Kondo.
\newblock {\em Molecular theory of surface tension in liquids}.
\newblock Springer - Verlag, Inc., Berlin - Gettingen - Heidelberg, 1960.

\bibitem{hillstatmeh1987}
T.~L. Hill.
\newblock {\em Statistical Mechanics: Principles and selected applications}.
\newblock Dover Publications, Inc., New York, 1987.

\bibitem{Bellemans1962}
A.~Bellemans.
\newblock Statistical mechanics of surface phenomena.
\newblock {\em Physica}, 28:493--510, 1962.

\bibitem{Bakri1966}
M.~M. Bakri.
\newblock Effect of surface tension on multilayer gas adsorption at moderate
  pressure.
\newblock {\em J. Chem. Phys.}, 44:2488--2495, 1966.

\bibitem{SteckiSokolowski1978}
J.~Stecki and S.~Sokolowski.
\newblock Fourth virial coefficient for a hard-sphere gas interacting with a
  hard wall.
\newblock {\em Phys. Rev. A}, 18(5):2361--2365, 1978.

\bibitem{SokolowskiStecki1980}
J.~Stecki and S.~Sokolowski.
\newblock The surface second virial coefficient.
\newblock {\em Mol. Phys.}, 39:343--351, 1980.

\bibitem{SokolowskiStecki1981}
S.~Sokolowski and Stecki J.
\newblock Second surface virial coefficient for argon adsorbed on graphite.
\newblock {\em J. Phys. Chem.}, 85:1741--1746, 1981.

\bibitem{LairdDavidchak2010}
B.~B. Laird and R.~L. Davidchak.
\newblock Calculation of the interfacial free energy of a fluid at a static
  wall by gibbs-cahn integration.
\newblock {\em J. Chem. Phys}, 132:204101, 2010.

\bibitem{IrvingKirkwood1950}
J.~H. Irving and J.~G. Kirkwood.
\newblock The statistical mechanical theory of transport processes. iv. the
  equations of hydrodynamics.
\newblock {\em J. Chem. Phys}, 18(6):817--829, 1950.

\bibitem{Harasima1958}
A.~Harasima.
\newblock Molecular theory of surface tension.
\newblock {\em Adv. Chem. Phys.}, 1:203--237, 1958.

\bibitem{HendersonSwol1984}
J.~R. Henderson and F.~van Swol.
\newblock On the interface between a fluid and a planar wall. theory and
  simulations of a hard sphere fluid at a hard wall.
\newblock {\em Mol. Phys.}, 51:991--1010, 1984.

\bibitem{Navascues1979}
G.~Navascues.
\newblock Liquid surfaces: Theory of surface tension.
\newblock {\em Rep. Prog. Phys}, 42:1131--1186, 1979.

\bibitem{NavascuesBerry1977}
G.~Navascues and M.~V. Berry.
\newblock The statistical mechanics of wetting.
\newblock {\em Mol. Phys}, 34(3):649--664, 1977.

\bibitem{StillingerBuff1962}
F.~H. Stillinger and F.~P. Buff.
\newblock Equilibrium statistical mechanics of inhomogeneous fluids.
\newblock {\em J. Chem. Phys}, 37(1):1--12, 1962.

\bibitem{SchofieldHenderson1982}
P.~Schofield and J.~R. Henderson.
\newblock Statistical mechanics of inhomogeneous fluids.
\newblock {\em Proc. R. Soc. Lond. A}, 379:231--246, 1982.

\bibitem{KirkwoodBuff1949}
J.~G. Kirkwood and F.~P. Buff.
\newblock The statistical mechanical theory of surface tension.
\newblock {\em J. Chem. Phys}, 17(3):338--343, 1949.

\bibitem{BrykPatrykiejewSokolowski2000}
P.~Bryk, A.~Patrykiejew, and S.~Sokolowski.
\newblock Surface phase transitions of a lennard-jones fluid in contact with a
  permeable wall of finite thickness : a density functional approach.
\newblock {\em Phys. Chem. Chem. Phys.}, 2:3227--3234, 2000.

\bibitem{Zaskulnikov201102a}
V.~M. Zaskulnikov.
\newblock Statistical mechanics of fluids at an impermeable wall.
\newblock 2011,  \href{http://arxiv.org/abs/1005.1063}{arXiv:1005.1063
  [cond-mat.stat-mech]}.

\bibitem{landaulifshitz1985}
L.~D. Landau and E.~M. Lifshitz.
\newblock {\em Statistical Physics}, volume~5.
\newblock Pergamon Press, Oxford - New York - Toronto - Sydney - Paris -
  Frankfurt, 3 edition, 1985.

\bibitem{percus1964}
J.~K. Percus.
\newblock The pair distribution function in classical statistical mechanics.
\newblock In H.~L. Fricsh and H.~D. Lebowitz, editors, {\em The equilibrium
  theory of classical fluids}, pages II--33 -- II--170. W. A. Benjamin, inc.,
  New York Amsterdam, 1964.

\bibitem{UhlenbeckFord1962}
G.~E. Uhlenbeck and G.~W. Ford.
\newblock The theory of linear graphs with applications to the theory of the
  virial development of the properties of gases.
\newblock In J.~de~Boer and G.~E. Uhlenbeck, editors, {\em Studies in
  statistical mechanics}, volume~1, pages 119 -- 207. North-Holland Publishing
  Co, Amsterdam, 1962.

\bibitem{zaskulnikov200911a}
V.~M. Zaskulnikov.
\newblock Open statistical ensemble and surface phenomena.
\newblock 2009,  \href{http://arxiv.org/abs/0911.3106}{arXiv:0911.3106
  [cond-mat.stat-mech]}.

\bibitem{RowlinsonWidom2002}
J.~S. Rowlinson and B.~Widom.
\newblock {\em Molecular theory of capillarity}.
\newblock Dover Publications, Inc., New York, 2002.

\bibitem{RuelleStatmeh1969}
D.~Ruelle.
\newblock {\em Statistical Mechanics. Rigorous Results}.
\newblock W. A. Benjamin, Inc., New York - Amsterdam, 1969.

\bibitem{zaskulnikov201004a}
V.~M. Zaskulnikov.
\newblock Open statistical ensemble: new properties (scale invariance,
  application to small systems, meaning of surface particles, etc.).
\newblock 2010,  \href{http://arxiv.org/abs/1004.0896}{arXiv:1004.0896
  [cond-mat.stat-mech]}.

\end{thebibliography}
\end{document}